\documentclass[fleqn,usenatbib]{mnras}

\usepackage{newtxtext,newtxmath}

\usepackage[T1]{fontenc}

\DeclareRobustCommand{\VAN}[3]{#2}
\let\VANthebibliography\thebibliography
\def\thebibliography{\DeclareRobustCommand{\VAN}[3]{##3}\VANthebibliography}

\usepackage{graphicx}	
\usepackage{amsmath}	
\usepackage{pdflscape}
\usepackage{afterpage}

\newcommand{\xmm}{\textit{XMM-Newton}}

\newcommand{\erosita}{{\it eROSITA}}

\newcommand{\fvar}{$F_\mathrm{var}$}

\title[X-ray Variability in 1H~0707-495]{The nature of the extreme X-ray variability in the NLS1 1H~0707-495}

\author[M. L. Parker et al.]{M. L. Parker,$^{1}$\thanks{E-mail: mlparker@ast.cam.ac.uk}
W. N. Alston,$^{2}$\thanks{E-mail: walston@sciops.esa.int}
L. H\"arer,$^{3}$
Z. Igo,$^{2}$
A. Joyce,$^3$
D. J. K. Buisson,$^4$\newauthor
P. Chainakun,$^5$
A. C. Fabian,$^1$
J. Jiang,$^6$
P. Kosec,$^7$
G. A. Matzeu,$^{8,2}$
C. Pinto,$^9$
Y. Xu$^{9,10}$\newauthor
F. Zaidouni$^1$
\\
$^{1}$Institute of Astronomy, Madingley Road, Cambridge, CB3 0HA, UK\\
$^{2}$European Space Agency (ESA), European Space Astronomy Centre (ESAC), E-28691 Villanueva de la Ca\~{n}ada, Madrid, Spain\\
$^{3}$Dr. Karl Remeis-Observatory \& ECAP, University of Erlangen-Nuremberg,    Sternwartstr. 7, 96049 Bamberg, Germany\\
$^4$Department of Physics and Astronomy, University of Southampton, Highfield, Southampton, SO17 1BJ\\
$^5$School of Physics, Institute of Science, Suranaree University of Technology, Nakhon Ratchasima, 30000, Thailand\\
$^6$Department of Astronomy, Tsinghua Univerisity, Shuangqing Road, Beĳing 100084, China\\
$^7$MIT Kavli Institute for Astrophysics and Space Research, Cambridge, MA 02139, USA\\
$^8$Department of Physics and Astronomy (DIFA), University of Bologna, Via Gobetti, 93/2, I-40129 Bologna, Italy\\
$^9$INAF - IASF Palermo, Via U. La Malfa 153, I-90146 Palermo, Italy\\
$^{10}$Università degli Studi di Palermo, Dipartimento di Fisica e Chimica, via Archirafi 36, I-90123 Palermo, Italy
}

\date{Accepted XXX. Received YYY; in original form ZZZ}

\pubyear{2020}

\begin{document}
\label{firstpage}
\pagerange{\pageref{firstpage}--\pageref{lastpage}}
\maketitle
\begin{abstract}
We examine archival \xmm\ data on the extremely variable narrow-line Seyfert 1 (NLS1) active galactic nucleus (AGN) 1H~0707-495. We construct fractional excess variance ($F_\mathrm{var}$) spectra for each epoch, including the recent 2019 observation taken simultaneously with \erosita . We explore both intrinsic and environmental absorption origins for the variability in different epochs, and examine the effect of the photoionised emission lines from outflowing gas.
In particular, we show that the unusual soft variability first detected by \erosita\ in 2019 is due to a combination of an obscuration event and strong suppression of the variance at 1~keV by photoionised emission, which makes the variance below 1~keV appear more extreme.
We also examine the variability on long timescales, between observations, and find that it is well described by a combination of intrinsic variability and absorption variability. We suggest that the typical extreme high frequency variability which 1H~0707-495 is known for is intrinsic to the source, but the large amplitude, low frequency variability that causes prolonged low-flux intervals is likely dominated by variable low-ionisation, low velocity absorption.

\end{abstract}

\begin{keywords}
galaxies: active -- accretion, accretion disks -- black hole physics
\end{keywords}


\section{Introduction}
Variability in flux and spectral shape gives an additional dimension to studies of compact objects, which in general cannot be spatially resolved. This variability can be broadly divided into two categories: variability which is intrinsic to the accretion flow, where the emitted flux from the accretion disk and corona changes; and variability caused by absorption and scattering along the line of sight to the AGN changing the observed flux.
Both of these phenomena are of scientific interest. Absorption variability typically tells us about the environment of the AGN, such as clumps of material in the broad line region \citep[BLR; e.g.][]{Gallo21}. Intrinsic variability, on the other hand, tells us about the accretion disk and corona on small scales \citep[e.g.][]{Uttley14, Alston20_natast}. The ultra-fast outflows (UFOs), winds launched from the accretion disk close to the black hole \citep[e.g.][]{Pounds03}, sit somewhere in between these two categories. Their main observational characteristic is high ionisation absorption lines in the X-ray spectrum, but the variability of these features appears to be driven by X-ray continuum \citep[e.g.][Xu et al., in prep]{Parker17_nature,Matzeu17,Pinto18} and in observational terms can be thought of as an enhancement to the intrinsic variability \citep[][]{Parker20}.

To use variability to study AGN, it is therefore vital that we understand what is driving the observed variability in each case. Arguably the simplest way to do this is by measuring the energy dependence of the variability, as different mechanisms predict different variability patterns \citep[e.g.][]{Parker15_pcasample}. Recently, we showed that the absorption lines from UFOs respond to the X-ray continuum flux \citep[][]{Parker17_nature, Matzeu17}, enhancing the variability in these energy bands, leading to spikes in variance spectra \citep[][]{Parker17_irasvariability, Parker18_pds456}. This gives us a new, complimentary tool for outflow detection \citep[][]{Igo20}, and a powerful probe of the physics involved. To exploit this fully, we need to be able to quantify the properties of the variability spectra. We have developed a scheme for modelling variance spectra within \textsc{xspec} based on simple, physically motivated, public models\footnote{\url{www.michaelparker.space/variance-models}}. We have used these to fit the variance spectra of IRAS~13224-3809 \citep[][]{Alston20_natast, Parker20} and PDS~456 \citep[][]{Haerer21}. 

Using these models, we were able to show that intrinsic variability, with damping where the soft excess and reflection spectrum are less variable and enhancement from UFO lines, can provide an excellent description of the variance spectrum of IRAS~13224-3809 \citep[][]{Alston20_natast,Parker20}. Building on this, we were able to recover the X-ray luminosity/UFO velocity correlation identified by \citet{Matzeu17} in PDS 456 by fitting the variance spectrum \citep[][]{Haerer21}. We were also able to detect three layers of UFO absorption with different velocities \citep[previously detected with different instruments in different observations:][]{Reeves09, Reeves16, Reeves18_pds456} in a single variance spectrum.

Similar to IRAS~13224-3809, 1H~0707-495 is a low redshift (z=0.04) NLS1 galaxy, well known for its extreme variability and spectral shape \citep[e.g.][]{Turner99, Leighly99, Boller02}, with a very strong soft excess and relativistic broad iron line from emission reprocessed by the accretion disk \citep{Fabian09}. NLS1s are generally thought to be low mass, high accretion rate AGN \citep[see review by][]{Komossa08}, making them highly variable and also interesting test-cases for AGN feedback in the early universe.
1H~0707-495 was also the first AGN where an X-ray reverberation lag was detected \citep[][]{Fabian09}, which was later shown to have a strong iron line feature in the lag-energy spectrum \citep{Kara13_1h0707}.
In addition, blueshifted absorption features from an ultra-fast outflow (UFO) are present in the \xmm\ spectra of 1H~0707-495 \citep[][]{Dauser12, Hagino16}, and emission lines from O\textsc{viii} and N\textsc{vii} are present in the reflection grating spectrometer (RGS) spectrum \citep[][]{Kosec18} when the continuum flux is low. Various authors have at times invoked low ionization partial-covering absorption in 1H~0707-495, either as a way of producing spectral structure at 7~keV at the Fe~K edge \citep[e.g.][]{Mizumoto14}, or as a way of producing soft X-ray variability \citep[e.g.][]{Boller21}.

In this paper, we use variance spectroscopy to examine the variability behaviour of each epoch of observations of 1H~0707-495 with \xmm , aiming to determine the physical mechanism(s) driving its variability. 1H 0707-495 is one of the most observed AGN in X-rays, with a total exposure duration of $\sim 1.4$\,Ms, and its extreme variability make it an ideal candidate for this kind of study.

\section{Observations and Data Reduction}

1H 0707-495 has been observed with \xmm\ 16 times over a period of 20 years.
In this paper, we use data from the EPIC-pn camera \citep{struder01} only, due to its higher throughput.  The raw data were processed from Observation Data Files (ODFs) following standard procedures using the \xmm\ Science Analysis System (SAS; v19.0.0), using the filtering conditions {\tt PATTERN} 0--4 and {\tt FLAG} = 0.  The 2000 and 2002 EPIC observations were made using full-frame mode and the remainder are in large-window mode.  The source counts were extracted from a circular region with radius 20 arc seconds.  The background is extracted from a large rectangular region on the same chip and avoids the Cu ring on the outer parts of the pn chip.  No significant pile-up is found using the SAS task {\tt  EPATPLOT} (\citealt{ballet99}; \citealt{davis01}).

Strong background flares can affect timing results, which is often worse at higher energies, so particular care has been taken to remove the influence of flaring and background variations.  These are typically worse at the beginning and end of \xmm\ observations.  Following \citet{alston13b,alston14b,alston19a}, for flares of duration $\le 200$\,s the source light curve for the flare duration was removed and the gap was interpolated by adding Poisson noise using the mean of neighbouring points.  The interpolation fraction was typically $< 0.5$\,\%.  For gaps longer than $200$\,s the data were treated as separate segments.  Any segments where the background light curve was comparable to the source rate (in any energy band used) were excluded from the analysis. 

\begin{table}
\caption{\xmm\ observation properties.}
\centering
\begin{tabular}{lccccccc}
\hline
\hline
{\tt obsID}     & Rev. & Start Date & Duration &  Count rate   \\
          &          &            & [s]     & (0.5--10~keV) [s$^{-1}$]  \\
\hline
110890201 & 159      & 21/10/2000 & 46018    & $0.654\pm0.004$  \\
148010301 & 521      & 13/10/2002 & 79953    & $2.398\pm0.006$  \\
506200301 & 1360     & 14/05/2007 & 40953    & $1.175\pm0.006$  \\
506200201 & 1361     & 16/05/2007 & 40914    & $0.379\pm0.004$  \\
506200501 & 1379     & 20/06/2007 & 46913    & $3.63\pm0.01$  \\
506200401 & 1387     & 06/07/2007 & 42866    & $2.54\pm0.02$   \\
511580101 & 1491     & 29/01/2008 & 123815   & $2.162\pm0.004$ \\
511580201 & 1492     & 31/01/2008 & 123670   & $3.287\pm0.008$   \\
511580301 & 1493     & 02/02/2008 & 122504   & $2.759\pm0.007$  \\
511580401 & 1494     & 04/02/2008 & 121922   & $2.259\pm0.009$  \\
653510301 & 1971     & 13/09/2010 & 116575   & $2.396\pm0.005$  \\
653510401 & 1972     & 15/09/2010 & 128200   & $3.280\pm0.006$  \\
653510501 & 1973     & 17/09/2010 & 127602   & $2.328\pm0.005$ \\
653510601 & 1974     & 19/09/2010 & 129001   & $2.857\pm0.005$ \\
554710801 & 2032     & 12/01/2011 & 98321    & $0.175\pm0.002$ \\
853000101 & 3633     & 11/10/2019 & 60700    & $0.111\pm0.002$  \\
\hline
\hline
\end{tabular}
\end{table}

\begin{table*}
    \centering
    \caption{Breakdown of obsIDs in each epoch spectrum. Observations from the same observing campaign are grouped together into a single epoch unless they differ drastically in flux (as in the first two orbits in 2007). For each epoch, we report the low frequency and high frequency (LF and HF) $F_\mathrm{var}$ values, averaged over two different energy bands.}
    \label{tab:epoch_breakdown}
    
    \begin{tabular}{l c c c c c}
        \hline \hline
        
        Epoch & obsIDs & \multicolumn{4}{c}{$F_\mathrm{var}$}\\
         & & LF, 0.5--2 keV & LF, 2--10 keV & HF, 0.5--2 keV & HF, 2--10keV\\
        \hline
        2000 & 110890201 & $0.25\pm0.01$ & $0.25\pm0.05$ & $0.12\pm0.01$ & $0.24\pm0.04$\\
        2002 & 148010301 & $0.178\pm0.002$ & $0.21\pm0.02$ & $0.052\pm0.003$ & $0.11\pm0.01$\\
        2007a & 506200301 & $0.280\pm0.005$ & $0.23\pm0.04$ & $0.232\pm0.005$ & $0.28\pm0.04$\\
        2007b & 506200201 & $0.56\pm0.01$ & $0.50\pm0.03$ & $0.39\pm0.01$ & $0.6\pm0.1$\\
        2007c & 506200501, 506200401 & $0.170\pm0.002$ & $0.21\pm0.01$ & $0.087\pm0.002$ & $0.13\pm0.01$\\
        2008 & 511580101, 511580201, 511580301, 511580401 & $0.167\pm0.001$ & $0.18\pm0.01$ & $0.050\pm0.001$ & $0.10\pm0.01$\\
        2010 & 653510301, 653510401, 653510501, 653510601 & $0.212\pm0.001$ & $0.30\pm0.01$ & $0.053\pm0.001$ & $0.14\pm0.02$\\
        2011 & 554710801 & $0.21\pm0.01$ & $0.41\pm0.06$ & $0.20\pm0.01$ & $0.22\pm0.02$\\
        2019 & 853000101 & $0.69\pm0.01$ & $0.43\pm0.04$ & $0.47\pm0.02$ & $0.36\pm0.08$\\
        High flux & All 2008 + 2010 obsIDs & $0.239\pm0.001$ & $0.26\pm0.01$ & $0.053\pm0.001$ & $0.11\pm0.01$\\
        Inter-obs & 110890201, 148010301, 506200301, 506200201, 506200501,  & $0.679\pm0.001^*$ & $0.53\pm0.01^*$ \\
        & 506200401, 511580101, 653510301, 554710801, 853000101\\
        \hline
        \hline
    \end{tabular}
    
    $^*$Note that the frequency band for the inter-observation variance spectrum is not the same as the low frequency band used for other epochs, as it includes variance on very long timescales (between observations) so these values are not directly comparable.
\end{table*}

\subsection{Variability spectra}

The energy dependence of the variability is investigated using rms-, or  frequency-resolved spectra (e.g. \citealt{Edelson2002}; \citealt{MarkowitzETAL03b}; \citealt{vaughan03a}, \citealt{alston15}).  We calculate the rms in a given energy band by integrating the noise subtracted power-spectrum (PSD), using an rms normalisation, over the frequency range of interest from $f_{\rm low}$ to $f_{\rm high}$.
This gives the rms spectrum in absolute units.  The fractional excess variance ($F_\mathrm{var}$, also known as the fractional rms spectrum or normalised excess variance) is obtained by dividing the rms spectra by the mean count rate in each energy band \citep[e.g.][]{Revnivtsev99}.  Following \citet{PoutanenETAL08} we calculate errors using Poisson statistics.  Energy bands are made sufficiently broad such that no time bins have zero counts.

We compute the rms spectra in two broad frequency bands;  one at low frequencies; $0.8 - 2.0 \times 10^{-4}$\,Hz (LF), and one at high frequency; $0.4 - 1.0 \times 10^{-3}$\,Hz (HF).  These bands are chosen to be below and above the break frequency respectively (see e.g. \citealt{mchardy06, gonzalezmartin12}).  A full exploration of the energy dependent PSD will be shown in Alston et al, {\it in prep}.

We divide the data into different epochs, grouping obsIDs from continuous campaigns unless the count rates differ drastically between observations, and calculate average variance spectra for each epoch. A summary of the different epochs used in this work is presented in Table~\ref{tab:epoch_breakdown}. For simplicity, we only use the high signal EPIC-pn data. While including the MOS data would marginally improve the total counts, the improvement at high energies is negligible due to the lower effective area of the MOS detectors and the higher background contamination. Including the additional detectors would complicate the analysis, and risk introducing additional systematic effects and background contamination.

In some low-flux epochs, the time-average spectra of 1H~0707-495 are background dominated, due to its soft spectrum. In general, this is not a major problem for the reliability of an $F_\mathrm{var}$ analysis, so long as the background is stable and subtracted appropriately (see above). The main effect of a high background is to lead to undefined bins (where the predicted white noise variance exceeds the observed variance, and the square root in the RMS would return a negative value) and bins with very large errors. Both of these are visible in our low flux epochs (Section~\ref{sec:lowflux}, Fig.~\ref{fig:lowflux_spectra}), but neither of these will have a significant effect on the fits. In principle, this effect can result in a bias towards higher values of $F_\mathrm{var}$ \citep[see e.g.][]{WilkinsonUttley09},  however, this bias is only significant with a large number of bins, as the errors on the positive bins are very large so the impact they have on the fit is negligible. In the case of our low flux epochs, we only have two bins above 8~keV that are strongly affected, so the impact of this bias is negligible.

If the background is variable this could cause additional systematic effects if not properly subtracted. However, this is unlikely after removing flaring periods. We test this by examining the power spectra with and without background subtraction, and find minimal difference in the PSD shape. We conclude that the background is not significantly affecting the observed variability.

We also calculate a long-term variance spectrum by including inter-observation variability, allowing us to study the equivalent very-low frequency variability.  To ensure the more densely spaced observations (e.g. 2008 and 2010) do not dominate the resulting variance spectra, we choose just one observation from these years: obsIDs 511580101 and 653510301.  The gaps in the light curves between observations are then zero padded and the variance spectra are determined in a similar way to before.  The small difference in the shape of the energy dependent PSD (e.g. \citealt{gonzalezmartin12}) means the introduction of any energy-dependent aliasing effects will be negligible.  The resultant frequency range for the very-low frequency variance spectra is $0.01 - 1.0 \times 10^{-6}$\,Hz (VLF).

\section{New variance models}

We introduce here two new variance models, constructed as in \citet{Parker20}, which we use to model the variance spectra of 1H~0707-495. Both models are available from \url{www.michaelparker.space/variance-models}.

\subsection{Photoionised emission}
\label{sec:pion}
In addition to confirming the presence of a UFO in absorption in 1H~0707-495, \citet{Kosec18} detected blueshifted emission lines from N\textsc{vii} and O\textsc{viii} with a velocity of $\sim8000$~km~s$^{-1}$, which were interpreted as emission from an outflow on larger scales, driven by the UFO. 

Regardless of their physical origin, the presence of these ionised emission lines will have an effect on the observed variability of 1H~0707-495. As they are not relativistic, they must originate from a larger scale than the X-ray continuum and reflection spectrum. This means that the emission lines will be much less variable than the continuum, lowering the fractional variance in bands where emission lines are present.

For this model, we follow the same procedure as outlined in \citet{Parker20}. For a grid of values of ionization $\xi_\mathrm{PI}$, and flux fraction $f_\mathrm{PI}$ (defined as the fraction of the flux from 0.5--10 keV in the photoionised component in the log-averaged spectrum)\footnote{$f_\mathrm{PI}$ is defined in this way because the variance is calculated assuming a log-normal distribution of fluxes, however this definition will lead to values much larger than will be typically measured by fitting a time averaged spectrum, as that approach calculates fluxes based on the linear average and therefore assigns much more flux to the highly variable powerlaw component. The ratio between our parameter $f_\mathrm{PI}$ and the linear average is a function of the variance in the continuum, with the difference becoming more extreme as the variance increases.}, we vary a powerlaw spectrum in the presence of a constant photoionized emission component. We model the photoionized emission using an \textsc{xspec} table model version\footnote{\url{https://www.michaelparker.space/xspec-models}} of the \textsc{pion} model \citep{Miller15_nature, Mehdipour16} from \textsc{spex} \citep{Kaastra96}, as in \citet{Parker19_mrk335}, assuming a $\Gamma=2$ powerlaw spectral energy distribution (SED)\footnote{This is necessarily approximate, as the model is intended to be general rather than specific to a given source. The overall SED is unlikely to vary significantly on the timescales we study, so the adoption of a constant and approximate SED is unlikely to significantly affect our conclusions, beyond a small bias in ionisation and column density.}. At each grid point we calculate 1000 model flux spectra then calculate a variance spectrum from them, generating a variance spectrum at each grid point. We then divide these spectra by a reference powerlaw variance spectrum without photoionised emission, to isolate the effect that the emission has on the variance. We then convert this into a multiplicative table model in \textsc{xspec} format, which can be applied to an arbitrary continuum.

\begin{figure}
    \centering
    \includegraphics[width=\linewidth]{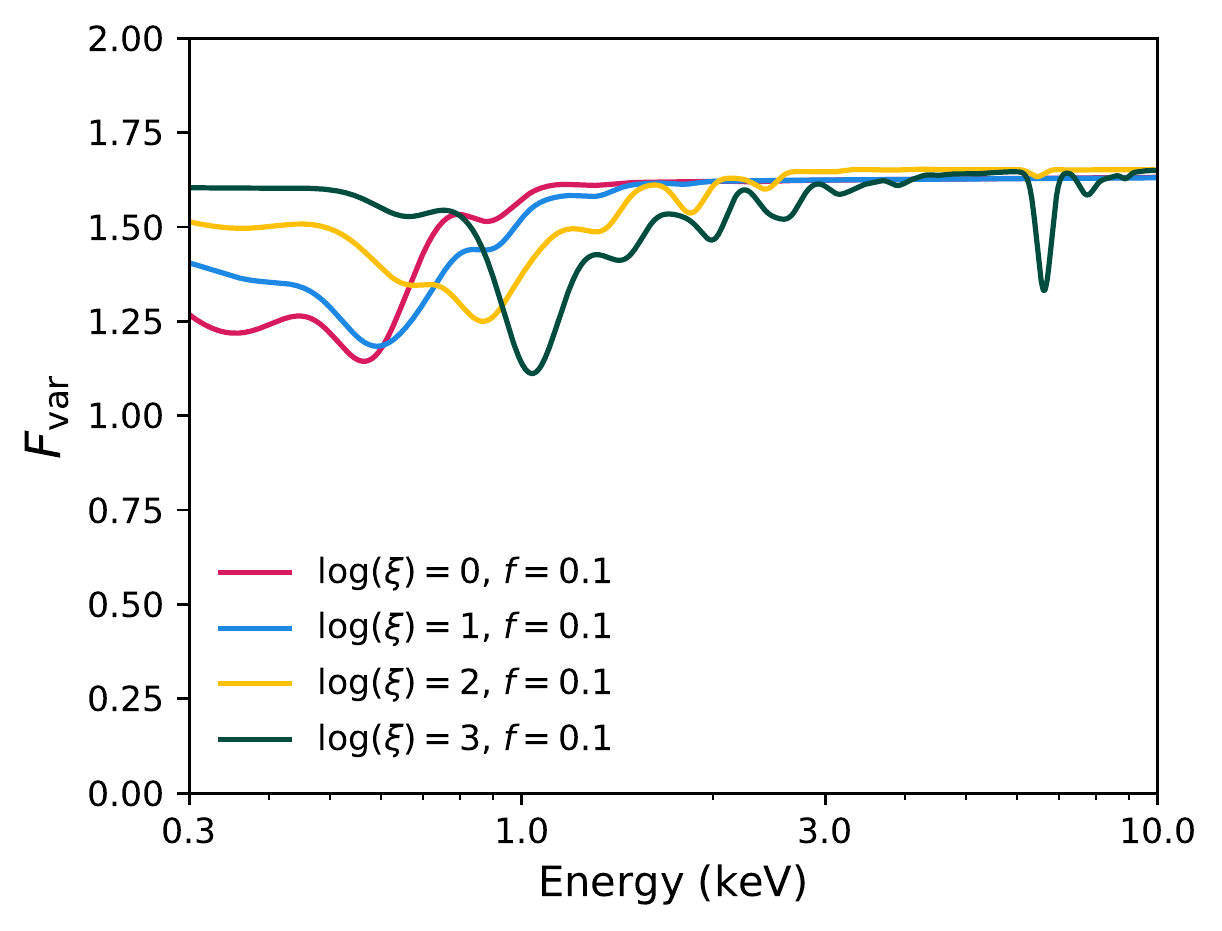}
    \caption{Variance spectra showing the effect of the photoionisation damping model \textsc{fvar\_pion}. Models are calculated for a variable powerlaw that varies in normalization only (no spectral pivoting) in the presence of constant photoionised emission (\textsc{fvar\_pion $\times$ fvar\_pow} in \textsc{xspec}), for four different values of the ionisation parameter. Spectra are smoothed to the approximate resolution of the EPIC-pn. A strong damping feature is present around 0.9--1~keV for values of $\log(\xi)$ between 2 and 3. At high ionisations a Fe line feature appears below 7~keV.}
    \label{fig:pionmodel}
\end{figure}

The main effect of this model is to lower the variability in energy bands where strong emission lines are predicted, which varies as a function of ionization. We plot this model, referred to in \textsc{xspec} as \textsc{pidamp}, for four values of ionisation in Fig.~\ref{fig:pionmodel}. Of particular relevance for this work are the spectra for $\log(\xi)=2$ and 3, where a strong dip is present around 1~keV. A similar dip is present in the observed spectra at low fluxes (see Section~\ref{sec:epochs})

\begin{figure}
    \centering
    \includegraphics[width=\linewidth]{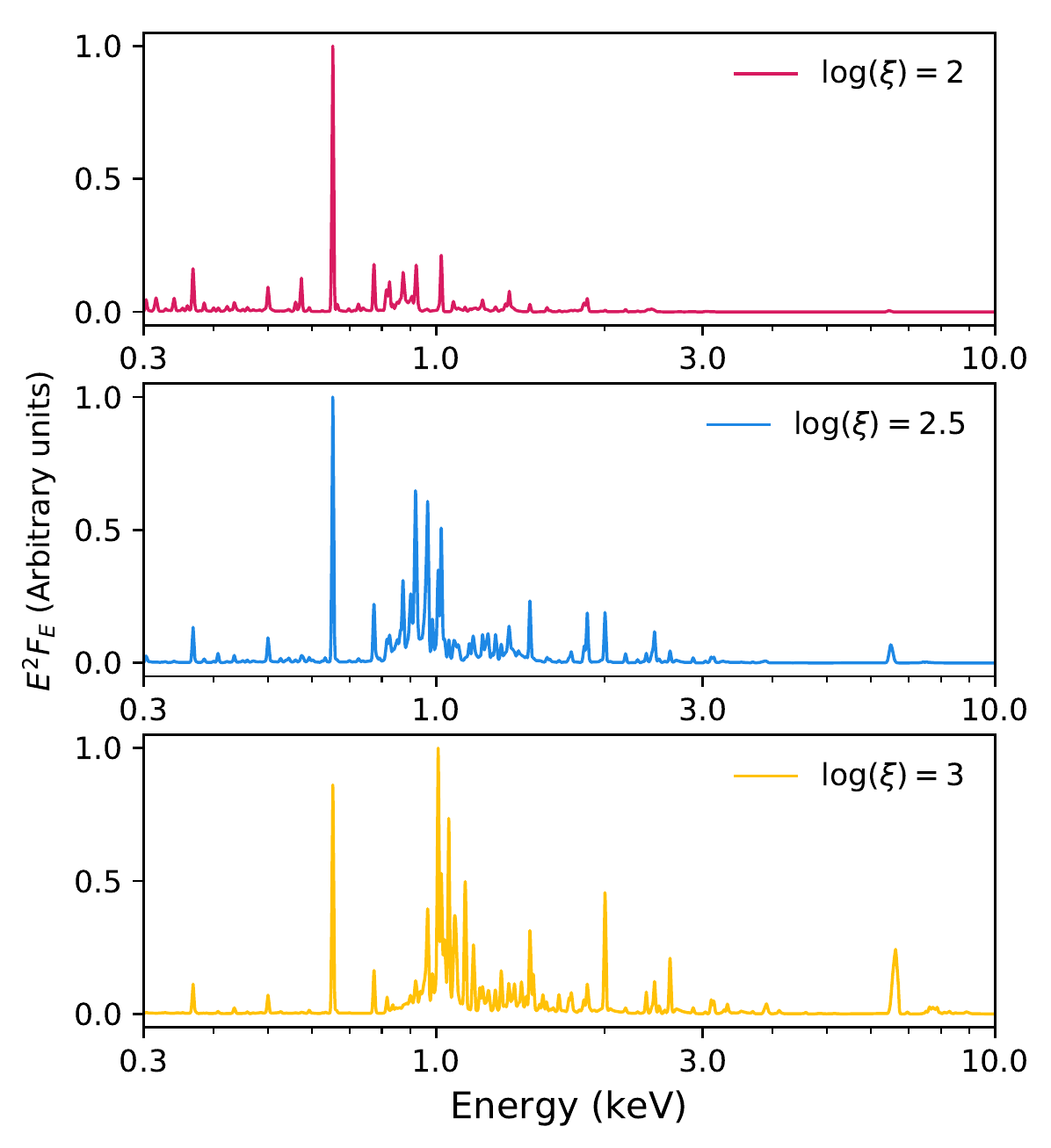}
    \caption{Emission spectra for the \textsc{pion} photoionised emission model, covering the range $\log(\xi)=2-3$. For values above $\sim2.3$ a strong forest of lines is present at 0.9--1~keV, shifting higher in energy as the ionization rises. A strong O\textsc{viii} line is present at 0.6~keV in all cases, which generally does not have a large effect on the variance spectra due to the limited energy resolution of the EPIC-pn.}
    \label{fig:pionmodel_emiss}
\end{figure}

We show corresponding emission spectra for this region of parameter space in Fig.~\ref{fig:pionmodel_emiss}. A forest of emission lines from various species of Ne and Fe is present around 1~keV, which causes the fractional variance to drop.

\subsection{Partial-covering absorption}

We construct a variance model for ionized partial covering absorption where the variability is driven by changes in covering fraction, with a potential correlation between the covering fraction and the column density.
A closely related model for neutral partial covering variance will be presented in Joyce et al. (in prep), where we use it to study variable absorption in a moderately obscured AGN.

\begin{figure}
    \centering
    \includegraphics[width=\linewidth]{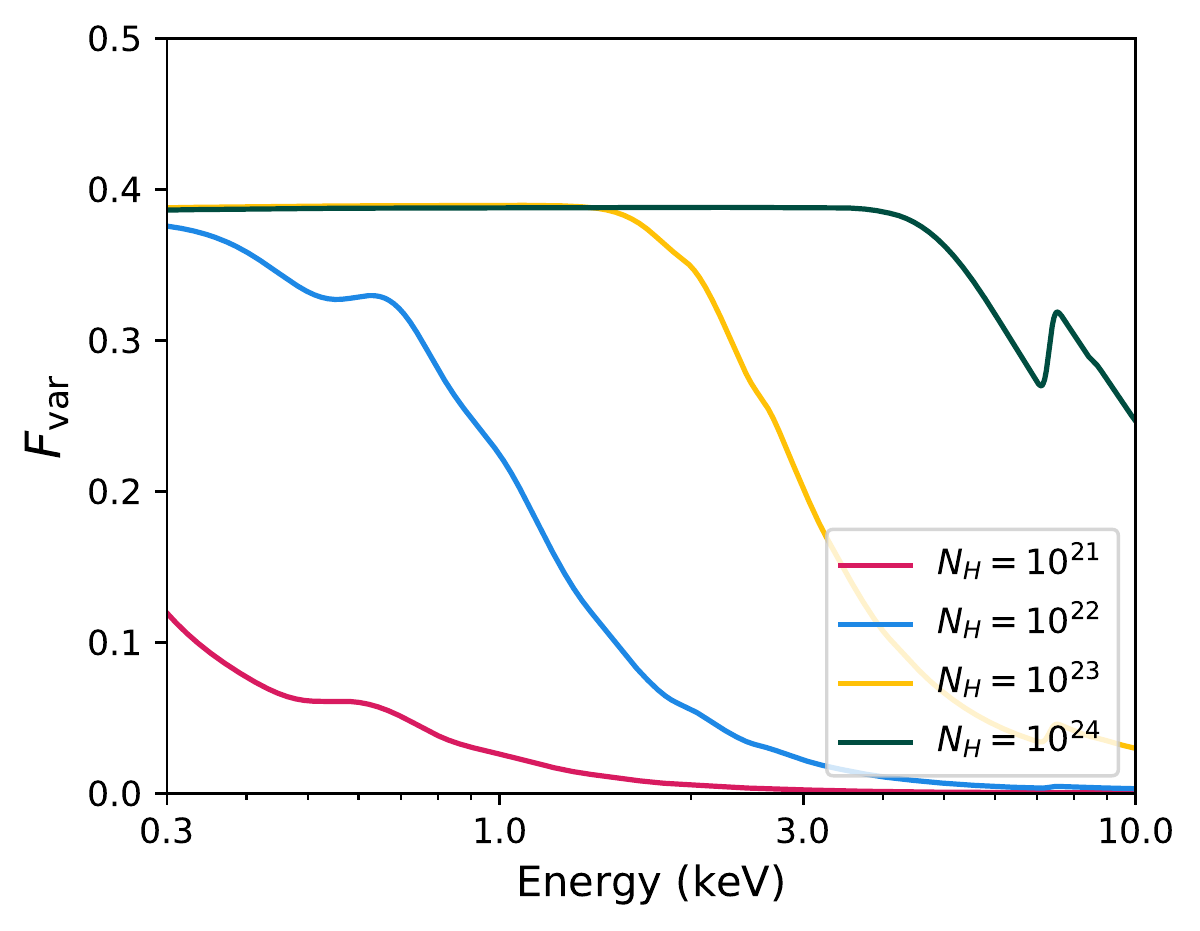}
    \includegraphics[width=\linewidth]{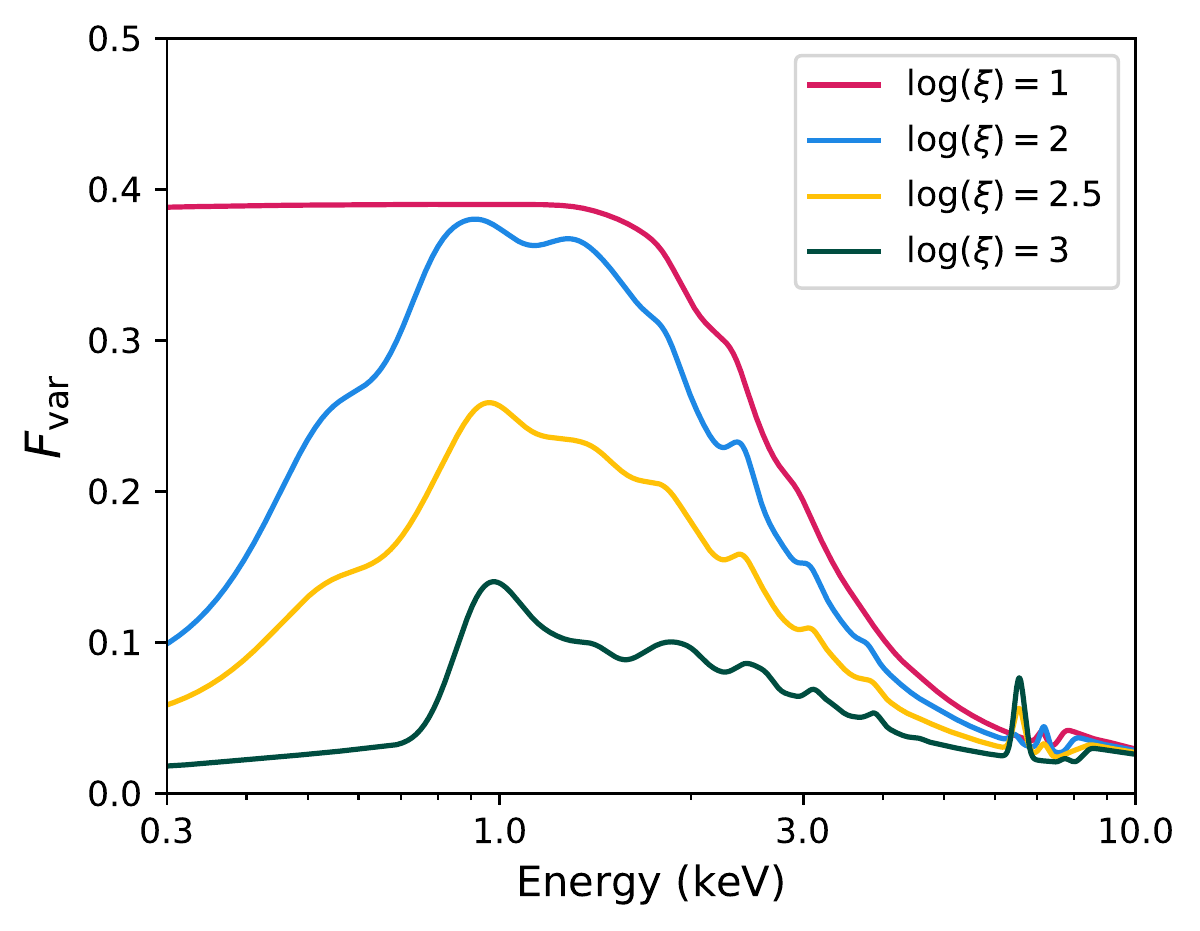}
    \caption{$F_\mathrm{var}$ spectra from our absorption variance model, \textsc{fvar\_pcov}, where the variability is driven by changes in the covering fraction. Top: Low ionisation ($\log(\xi)=0$) variance spectra, with different values of the column density. Bottom: Fixed column density ($N_\mathrm{H}=10^{23}$~cm$^{-2}$) spectra, with different values of the ionisation. In both cases we assume an average covering fraction $f_\mathrm{cov}$ of 0.1 and standard deviations in covering fraction $\sigma_{f_\mathrm{cov}}$ of 0.5. Note that as the ionization rises the variance below 1~keV drops, so ionised absorption does not provide an explanation for the extremely soft variability seen in 2019 (see Section~\ref{sec:2019}).}
    \label{fig:pcov_models}
\end{figure}

We use the same basic process as in the previous section to generate the model. We use the \textsc{xabs} model from \textsc{spex} to model the effect of variable absorption, again assuming a $\Gamma=2$ powerlaw SED. For a grid of values of average covering fraction, average column density, and ionization, we apply the absorption model to a powerlaw continuum. We then draw 1000 values of covering fraction from a Gaussian distribution centered on the average to calculate a range of spectra. If the instantaneous covering fraction would be outside the limits of 0--1 we pin it to the corresponding limit. From these spectra we calculate a variance spectrum for each grid point. 

We also account for the possibility that the column density correlates with covering fraction (for example, clouds are likely to be denser in the core, so when they fully cover the source the column would be higher) by including a correlation parameter, $c_{N_\mathrm{H}}$, with values between 0 and 1. This relates the instantaneous covering fraction and column density via the equation $N_\mathrm{H,i}=N_\mathrm{H,avg}\times(f_\mathrm{cov,i}/f_\mathrm{cov,avg})^{c_{N_\mathrm{H}}}$. Some example model spectra are shown in Fig.~\ref{fig:pcov_models}, for different values of ionisation and column density. We note that, regardless of ionization, none of these spectra show the sharp drop in variability above 1~keV observed in 1H~0707-495 by \citet[][their Fig. 3]{Boller21}, implying that absorption alone cannot fully explain the extremely soft variability spectrum.

\section{Results}

We now consider the variance spectra of 1H~0707-495 in different epochs and flux states. In each case, we attempt to model the spectrum at high ($0.4-1.0\times10^{-3}$~Hz, HF) and low ($0.8 - 2.0\times10^{-4}$~Hz, LF) frequencies simultaneously using our variance models, keeping most parameters fixed between the two frequencies and allowing only those that could be frequency dependent to differ between the two.

\begin{figure*}
    \centering
    \includegraphics[width=0.7\linewidth]{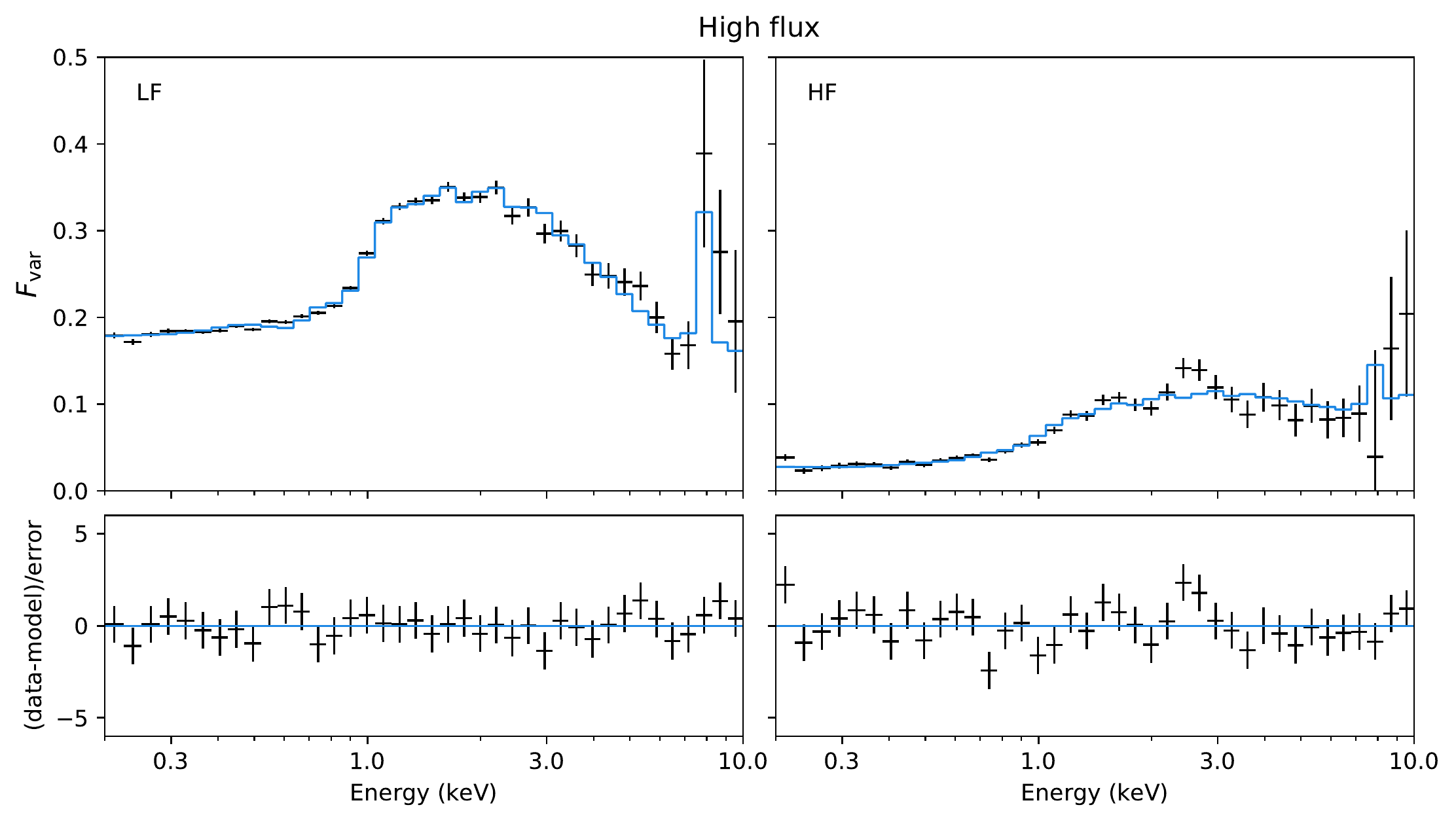}
    \caption{High flux variance spectrum, calculated from the long observing campaigns in 2008 and 2010, when the flux was relatively high. Low and high frequency spectra (LF and HF) are shown in the left and right panels, respectively. The best-fit model intrinsic variability model is plotted as a stepped blue line, in each case. For both LF and HF spectra, a strong damping effect from the soft excess is visible at low energies. A clear Fe UFO line is visible in the LF spectrum at 8~keV, and weaker lines above and below 2~keV from Si and S. The UFO lines are not clearly visible in the HF spectrum, but it is consistent with them still being present.}
    \label{fig:highflux}
\end{figure*}

\subsection{High flux average $F_\mathrm{var}$ spectrum}
\label{sec:highflux}

We first examine the combined 2008+2010 high flux spectrum, as this spectrum has the highest signal (and correspondingly highest energy resolution), allowing us to establish a baseline model.
The low-frequency spectrum (Fig.~\ref{fig:highflux}, left) is very similar to that of IRAS~13224-3809 \citep[][]{Parker20}, showing a depression of the variability below ~1.5 keV, corresponding to the soft excess, and a strong peak in variance at high energies corresponding to the Fe \textsc{xxv/xxvi} UFO line, with weaker peaks potentially corresponding to the equivalent lines for lower Z elements. The high-frequency spectrum is similar, but with a significantly lower power, making it harder to distinguish line features. The low energy dip is clearly present. 

It is interesting that the UFO variance declines in parallel with the continuum between the low and high frequency spectra. This means that it shares a common frequency dependence/power spectrum, consistent with the interpretation of the UFO variability being strongly correlated (and possibly driven by) the X-ray continuum. The effect of the UFO variability can therefore be regarded as an extension of the intrinsic source variability, rather than an independent source of variance.

The soft excess drop in variance is common in AGN \citep[e.g.][]{Igo20}, and is a result of the soft excess being less variable than the primary continuum, independently of the origin of the soft excess. It is important to note that neutral absorption variability does not produce an equivalent feature, as it affects both the power-law and soft excess photons equally. Ionized absorption can in principle produce a similar effect, as it preferentially absorbs photons at specific energies, however such absorption would produce clear atomic lines which would trivially be observed in grating spectra (and are not). The presence of this soft excess dip therefore argues strongly in favour of intrinsic variability dominating the variance spectrum.

We fit these spectra with the same intrinsic variability model used for IRAS~13224-3809 in \citet{Parker20}: a variable power-law, damped by a constant black-body soft excess and less variable relativistic reflection, and enhanced by a UFO that responds to the ionizing continuum. The powerlaw (\textsc{fvar\_pow}) produces the basic continuum shape, and the degree of pivoting of the powerlaw (defined by the correlation between photon index and flux, $c_\Gamma$) determines how soft it is. The relativistic reflection component (\textsc{fvar\_ref}) models the change in variance produced by a relativistic reflection spectrum \citep[modelled with \textsc{relxill,}][]{Garcia14}, where the flux in the reflection is correlated with the powerlaw ($\log(F_\mathrm{ref})=c_\mathrm{ref}\log(F_\mathrm{pow})+\log(F_\mathrm{ref,0})$). For values of $c_\mathrm{ref}<1$, the reflection spectrum is less variable than the powerlaw, and the reflection causes $F_\mathrm{var}$ to drop in bands where it is strong. The average reflection flux is set so that it is a fixed fraction of the average powerlaw flux, given by the $f_\mathrm{ref}$ parameter (note that this average reflection fraction is extremely degenerate with the reflection correlation, since low correlations and high fractions both produce strong damping, and vice-versa). We also include a black body damping model (\textsc{fvar\_bb}), which describes the damping effect of a constant black body component. We use this as a phenomenological soft excess component, in combination with the reflection model. The UFO model (\textsc{fvar\_ufo}) assumes that the variability in the UFO lines comes from changes in the ionisation of the UFO, driven by the powerlaw flux ($\log(\xi_\mathrm{UFO})=c_\mathrm{UFO}\log(F_\mathrm{pow})+\log(\xi_\mathrm{UFO,0})$). For any value of $c_\mathrm{UFO}>0$, the ionisation of the gas will increase with increasing powerlaw flux, and the strength of the absorption lines will change. For high ionization lines, such as the Lyman-$\alpha$ lines which typically dominate UFO spectra, this will cause the lines to weaken with increasing flux, enhancing the variance in that energy band.
For simplicity, as the UFO absorption variance model can easily become stuck in false minima, we fix the blueshift of the UFO to a velocity of 0.13$c$ \citep[][]{Kosec18}. While the velocity may vary between epochs \citep[][]{Dauser12}, the broad energy bins we use for the $F_\mathrm{var}$ spectra mean that the effect is minimal. In addition, we also add a constant photoionized emission component, which damps the variability at specific energies, to account for the emission seen by \citet{Kosec18}. The effect of this component is minimal in the high flux spectra, but we include it for comparison purposes. We convolve all these components with an energy-dependent Gaussian to mimic the instrumental resolution of the EPIC-pn, with the $\sigma$ at 6~keV fixed to 0.1~keV and the index of the energy dependence fixed to 0.165. In \textsc{xspec} notation, the model is: \textsc{gsmooth $\times$ (fvar\_pidamp $\times$ fvar\_ufo $\times$ fvar\_refdamp $\times$ fvar\_bbdamp $\times$ fvar\_pow)}.

The relativistic blurring parameters of the reflection component are fixed, as in \citet{Parker20}, to representative values for IRAS~13224-3809. This is unlikely to significantly affect the analysis, as the relativistic iron line has a negligible effect on the variance spectrum and the corresponding parameters for 1H~0707-495 are very similar.
Finally, we add a 2\% systematic error to all our fits to stop the fits being dominated by very high signal bins at low energies, where our relatively simplistic model (compared to those used to fit count spectra) may not perfectly describe the data. 

We fit the high and low frequency spectra simultaneously, and allow certain parameters which could be frequency-dependent to vary between the two. Specifically, we free the UFO ionization/power-law flux correlation, the reflection flux/power-law flux correlation, the power-law flux/index correlation, the black-body flux fraction (a proxy for how variable the soft excess is), and the power-law variance.

This model gives an excellent description of the data (Fig.~\ref{fig:highflux}), fitting all the main features (best-fit parameters in Table~\ref{tab:highres_bestfit}). A small residual is visible in the high-frequency spectrum around 2.5~keV, which is around the right energy for the blueshifted S\textsc{xvi} line \citep[e.g.][]{Dauser12}. However, we consider it unlikely that this is a genuine feature, as it is anomalously strong relative to the other UFO line signatures. There is sometimes an instrumental feature around this energy range \citep[see appendix of][]{Marinucci14}, and it is possible that this feature could be somehow variable and frequency dependent, introducing a spurious feature into the variance spectrum.

\begin{table}
    \centering
    \caption{Best-fit parameters for the high flux/high resolution spectrum shown in Fig.~\ref{fig:highflux}. Parameters with the subscripts PI, UFO, and Ref refer to the photoionised emission, UFO, and relativistic reflection, respectively. HF and LF refer to the high and low frequency spectra. Where only one value is given, parameters are tied between the two spectra.}
    \begin{tabular}{l c c r}
    \hline
    \hline
    Parameter & LF & HF & Description\\
    \hline
    
    $f_\mathrm{PI}$ & $ 0.14^{+0.01}_{-0.02} $ & & Photoionised fraction\\
    $\log(\xi_\mathrm{PI})$ & $ 2.1 \pm 0.03 $ & & Ionization (erg~cm~s$^{-1}$)\\
    $\log(\xi_\mathrm{UFO})$ & $ 4.7^{+0.1}_{-0.3} $ & & Ionization (erg~cm~s$^{-1}$)\\
    $N_\mathrm{H}$ & $ >6.8 $ & & Column density ($10^{24}$~cm$^{-2}$)\\
    $c_\mathrm{UFO}$ & $ 0.21^{+0.07}_{-0.1} $ & $ 0.33^{+0.01}_{-0.05} $ & Ionization correlation\\
    $n_\mathrm{Ref}$ & $ >18.97 $ & & Density (cm$^{-3}$)\\
    $f_\mathrm{Ref}$  & $ 0.55^{+0.08}_{-0.03} $ & & Reflected fraction\\
    $c_\mathrm{Ref}$ & $ 0.3^{+0.05}_{-0.1} $ & $ 0.4^{+0.3}_{-0.1} $ & Flux correlation\\
    $\log(\xi_\mathrm{Ref})$ & $ 1.67^{+0.1}_{-0.07} $ & & Ionization (erg~cm~s$^{-1}$)\\
    kT & $ 0.094^{+0.003}_{-0.002} $ & & Temperature (keV) \\
    $f_\mathrm{BB}$ & $ 0.39^{+0.02}_{-0.03} $ & $ 0.6^{+0.3}_{-0.1} $ & Black body fraction\\
    $\sigma$ & $ 0.171^{+0.008}_{-0.009} $ & $ 0.05^{+0.003}_{-0.006} $ & Powerlaw flux $\sigma$ \\
    $c_\Gamma$ & $ 0.72^{+0.1}_{-0.07} $ & $ >0.5 $ & Index correlation\\
    $\chi^2$/dof & 82/60 &  & Fit statistic\\
    
    \hline
    \hline
    \end{tabular}
    \label{tab:highres_bestfit}
\end{table}

\begin{landscape}
\begin{table}
    \caption{Best fit parameters for the fits to the low (LF) and high (HF) variance spectra in each epoch. Fluxes are calculated from phenomenological powerlaw plus black body fits to the count spectra, are in units of 10$^{-12}$ erg~cm$^{-2}$~s$^{-1}$, and are not corrected for Galactic absorption. Other parameters are defined as in Table~\ref{tab:highres_bestfit}.}
    \label{tab:epoch_fits}

\setlength{\tabcolsep}{2.5pt}
    \centering
    \begin{tabular}{l c c c c c c c c c c c c c c c c}
    \hline
    \hline
          &  \multicolumn{2}{c}{2000} & \multicolumn{2}{c}{2002} & \multicolumn{2}{c}{2007a} &  \multicolumn{2}{c}{2007b} &  \multicolumn{2}{c}{2007c} &  \multicolumn{2}{c}{2008} &  \multicolumn{2}{c}{2010} &  \multicolumn{2}{c}{2011} \\
         Par & LF & HF & LF & HF & LF & HF & LF & HF & LF & HF & LF & HF & LF & HF & LF & HF  \\
    \hline
 
        $f_\mathrm{PI}$ & $ 0.11^{+0.1}_{-0.06} $ &  & $ 0.05^{+0.03}_{-0.02} $ &  & $ 0.12^{+0.05}_{-0.04} $ &  & $ 0.23^{+0.08}_{-0.06} $ &  & $ 0.05 \pm 0.02 $ &  & $ 0.08 \pm 0.02 $ &  & $ 0.1 \pm 0.02 $ &  & $ 3.2^{+0.5}_{-0.3} $ &  \\

        $\log(\xi_\mathrm{PI})$ & $ >2.7 $ &  & $ 2.6^{+0.2}_{-0.5} $ &  & $ 2 \pm 0.2 $ &  & $ >3 $ &  & $ 2.6 \pm 0.2 $ &  & $ 2.19^{+0.08}_{-0.06} $ &  & $ 2.2 \pm 0.05 $ &  & $ 2.7^{+0.07}_{-0.06} $ &  \\

        $f_\mathrm{Ref}$ & $ >0.41 $ &  & $ 1.1^{+2}_{-0.6} $ &  & $ 0.5 \pm 0.08 $ &  & $ 0.6^{+0.3}_{-0.1} $ &  & $ 0.56^{+0.6}_{-0.06} $ &  & $ 0.9^{+0.6}_{-0.3} $ &  & $ 0.56^{+0.2}_{-0.05} $ &  & $ 0.5^{+0.4}_{-0.1} $ &  \\

        $c_\mathrm{Ref}$ & $ >0.54 $ &  & $ 0.34^{+0.5}_{-0.1} $ &  & $ >0.5 $ &  & $ <0.32 $ &  & $ <0.33 $ &  & $ 0.59^{+0.06}_{-0.09} $ &  & $ <0.2 $ &  & $ <0.38 $ &  \\

        $\log(\xi_\mathrm{Ref})$ & * &  & $ 2.4^{+0.2}_{-0.3} $ &  & $ 1.4^{+0.4}_{-0.3} $ &  & $ 2.39^{+0.07}_{-0.2} $ &  & $ 2.59^{+0.08}_{-0.1} $ &  & $ 2^{+0.2}_{-0.04} $ &  & $ 2.42^{+0.2}_{-0.08} $ &  & $ 2 \pm 0.1 $ &  \\

        $f_\mathrm{BB}$ & $ <0.14 $ &  & $ 0.3 \pm 0.09 $ &  & $ 0.5^{+0.2}_{-0.1} $ &  & $ 0.6^{+0.2}_{-0.1} $ &  & $ 0.36^{+0.08}_{-0.09} $ &  & $ 0.36^{+0.04}_{-0.05} $ &  & $ 0.44^{+0.05}_{-0.03} $ &  & $ 0.4^{+0.2}_{-0.1} $ &  \\

        $\sigma$ & $ 0.11^{+0.04}_{-0.01} $ & $ 0.08^{+0.02}_{-0.01} $ & $ 0.13^{+0.01}_{-0.02} $ & $ 0.059^{+0.006}_{-0.01} $ & $ 0.18 \pm 0.01 $ & $ 0.16 \pm 0.01 $ & $ 0.38^{+0.03}_{-0.02} $ & $ 0.28^{+0.02}_{-0.01} $ & $ 0.121^{+0.02}_{-0.005} $ & $ 0.074^{+0.009}_{-0.004} $ & $ 0.1 \pm 0.007 $ & $ 0.046^{+0.001}_{-0.003} $ & $ 0.18 \pm 0.008 $ & $ 0.063 \pm 0.004 $ & $ 0.33^{+0.04}_{-0.02} $ & $ 0.27^{+0.05}_{-0.03} $ \\

        $c_\Gamma$ & $ 0.6 \pm 0.2 $ & $ <0.18 $ & $ 0.8 \pm 0.1 $ & $ >0.5 $ & $ 0.9^{+0.3}_{-0.4} $ & $ 0.5 \pm 0.3 $ & $ 1.1^{+0.1}_{-0.2} $ & $ 0.7 \pm 0.1 $ & $ 0.6^{+0.1}_{-0.2} $ & $ 0.3^{+0.2}_{-0.1} $ & $ 0.83^{+0.06}_{-0.07} $ & $ >0.5 $ & $ 0.8 \pm 0.05 $ & $ >0.5 $ & $ 0.67^{+0.07}_{-0.09} $ & $ 1 \pm 0.2 $ \\
        
        $\chi^2$/dof & 31/28 & & 31/28 & & 49/30 & & 33/27 & & 29/27 & & 72/48 & & 81/49 & & 43/26 & \\
        
        $F_\mathrm{0.5-2}$ & 1.1 & & 4.0 & & 1.9 & & 0.61 & & 5.6  & & 4.3   & &4.6 & & 0.26\\

        $F_\mathrm{2-10}$ & 0.44 & & 1.1 & & 0.77 & & 0.31 & & 1.5 & & 1.2  & &0.93 & & 0.28\\
    
    \hline
    \hline
    
    \end{tabular}
\end{table}

\end{landscape}

We note that the column density of the UFO is extremely high. There are likely two reasons for this. Firstly, the main effect of this parameter is to scale the strength of the UFO features. This means that it is likely very degenerate with other parameters that have the same function, in this case the UFO ionisation/powerlaw flux correlation parameter $c_\mathrm{UFO}$. The value of $c_\mathrm{UFO}$ here is 0.2--0.3, which is much lower than the values of 0.6 found in IRAS~13224-3809 \citep[][]{Parker20} and 0.5--1 in PDS~456 \citep[][]{Haerer21}. Secondly, we note that the Fe UFO line is much stronger here than in the other two sources. $F_\mathrm{var}$ at the peak of the line is double the continuum on either side, whereas in the IRAS~13224-3809 and PDS~456 spectra the line peak is only 20--25\% higher than the continuum. It is likely that the UFO variance model cannot explain this high variance with the correlation parameter alone. Increasing the value of $c_\mathrm{UFO}$ does not just increase the variance of the lines, it also increases the range of ionisation values covered by the model, so it introduces strong lower energy line features from other ions, such as Si\textsc{xiv} and S\textsc{xvi}. These features are present in the $F_\mathrm{var}$ spectra of IRAS~13224-3809 and PDS~456, but only weakly in 1H~0707-495.

It is interesting that the parameters of the photoionised emission are well constrained, and broadly consistent with those found in the following epoch-resolved fits, despite there being no clear corresponding features in the variance spectrum. We speculate that this is because it subtly modifies the shape of the soft excess damping, leading to a better fit. 

\subsection{Epoch-resolved $F-\mathrm{var}$ spectra}
\label{sec:epochs}

We next fit each individual epoch with the same intrinsic variability model. Because the data quality is typically lower, we freeze some parameters to simplify the model. We fix all the parameters of the UFO, the reflection density, and the black body temperature to the best-fit values from the high resolution spectrum (Table~\ref{tab:highres_bestfit}). We note that a shift in the velocity of the UFO was observed between 2008 and 2010 by \citet{Dauser12}, however this is unlikely to have a major impact on our results with the limited signal and resolution of the epoch resolved spectra. The best fit parameters for each epoch are presented in Table~\ref{tab:epoch_fits}.

The individual epoch spectra can be broadly divided into three groups: those that come from high flux observations and are qualitatively very similar to the high flux/high resolution spectrum; those that come from low flux states, which are typically noisier, and show a strong damping feature around 1~keV; and the 2019 spectrum, which shows extreme soft variability.

\subsubsection{High flux: 2002, 2007a, 2007c, 2008, and 2010}

\begin{figure*}
    \centering
    \includegraphics[width=0.7\linewidth]{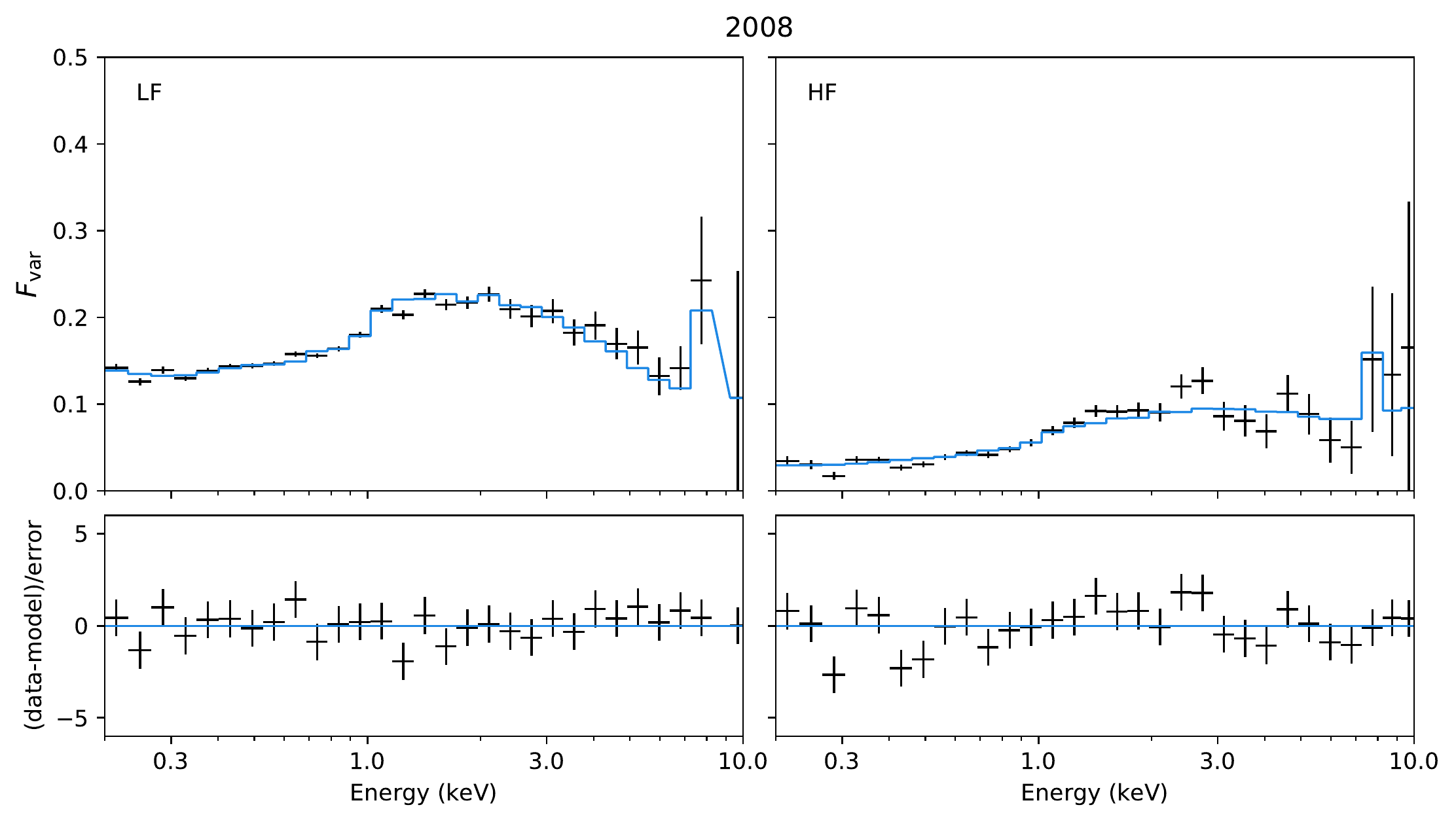}
    \includegraphics[width=0.7\linewidth]{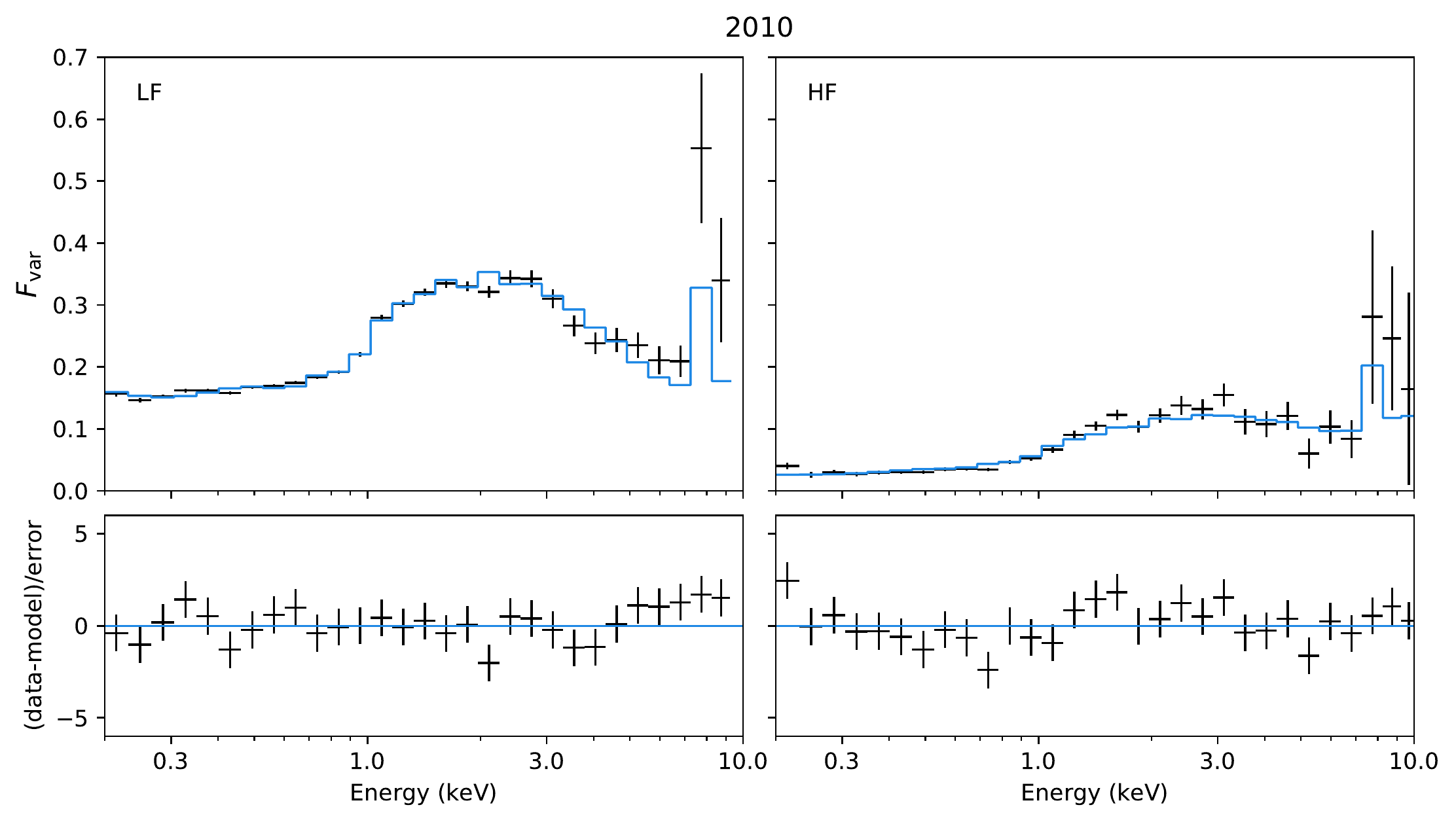}
    \caption{Fits to the 2008 and 2010 high-flux epochs. In each case the high and low frequency fits are shown in the left and right panels, respectively. Top panels show the data and model (blue stepped line). Bottom panels show the residuals in units of $\Delta\chi$.The 2008 and 2010 observing campaigns have higher signal than the other epochs due to longer exposures, and the Fe UFO line can be seen in the spectra. As with the high flux average spectrum (Fig.~\ref{fig:highflux}), the UFO line gets weaker with frequency, following the intrinsic variability of the source. The spectra for the shorter high flux exposures are shown in Fig.~\ref{fig:highflux}}
    \label{fig:highflux_spectra2}
\end{figure*}

The high flux epochs have relatively uniform spectra (Figs.~\ref{fig:highflux_spectra2},~\ref{fig:highflux_spectra}). They all show a dip at low energies due to damping from the soft excess (clear evidence that the variability in these observations is dominated by intrinsic variance), and a gradual decrease in variance at high energies, likely due to a combination of a larger contribution from less variable reflected emission at high energies and pivoting of the powerlaw (the softer when brighter behaviour commonly observed in AGN). Only the higher signal 2008 and 2010 observing campaigns show a strong iron absorption line variance spike at 8~keV, but the other spectra are consistent with the feature being present.

The fits are generally good, with only small residuals remaining. There are some possible features at intermediate energies in the 2007a spectrum, and the Fe UFO line is slightly too weak in the 2010 low frequency spectrum. We conclude that these spectra are well described by an intrinsic variability model, where the coronal variability dominates, damped by relativistic reflection and enhanced by the UFO.

\subsubsection{Low flux: 2000, 2007b, and 2011}
\label{sec:lowflux}
\begin{figure*}
    \centering
    \includegraphics[width=0.7\linewidth]{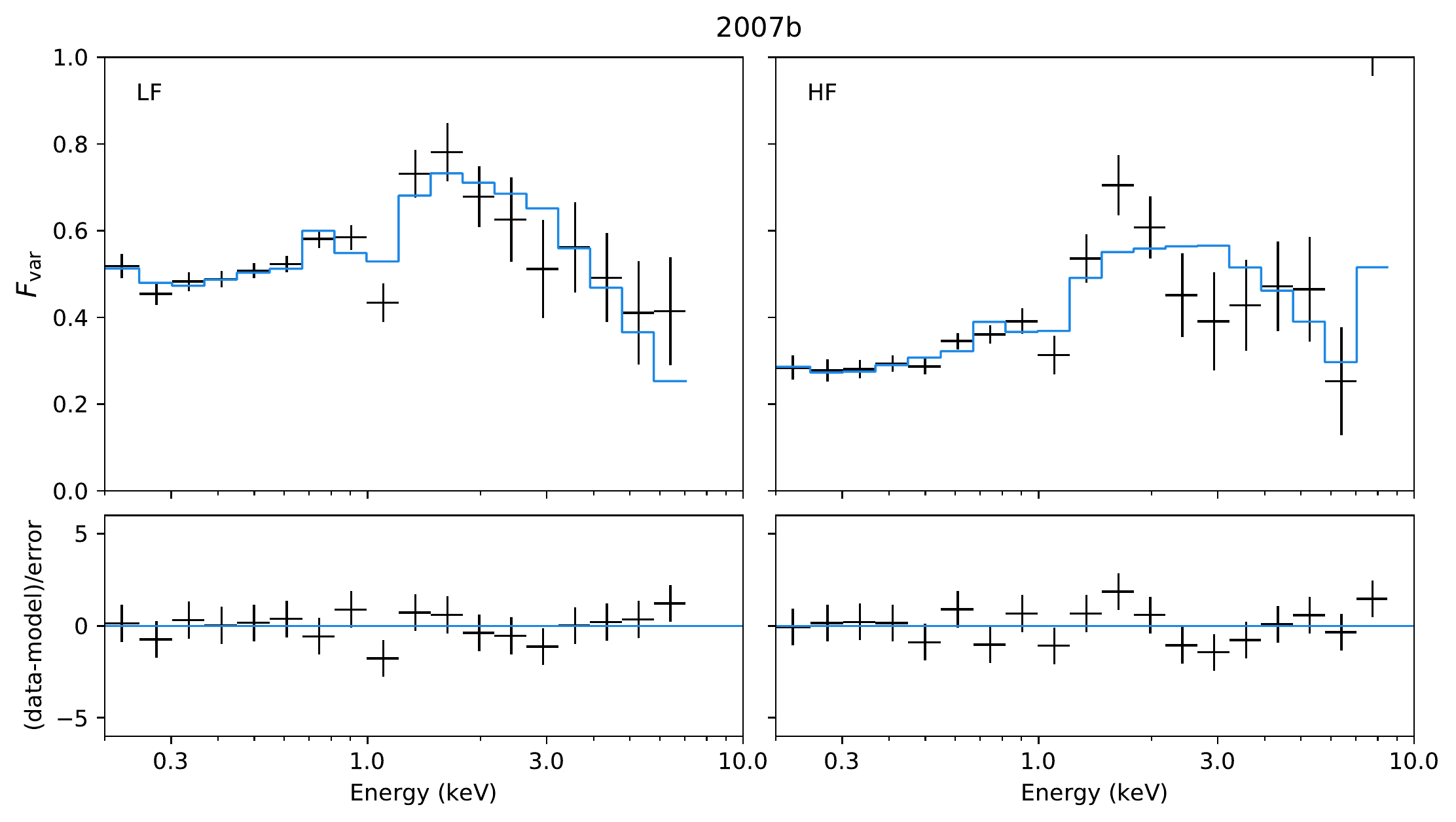}
    \includegraphics[width=0.7\linewidth]{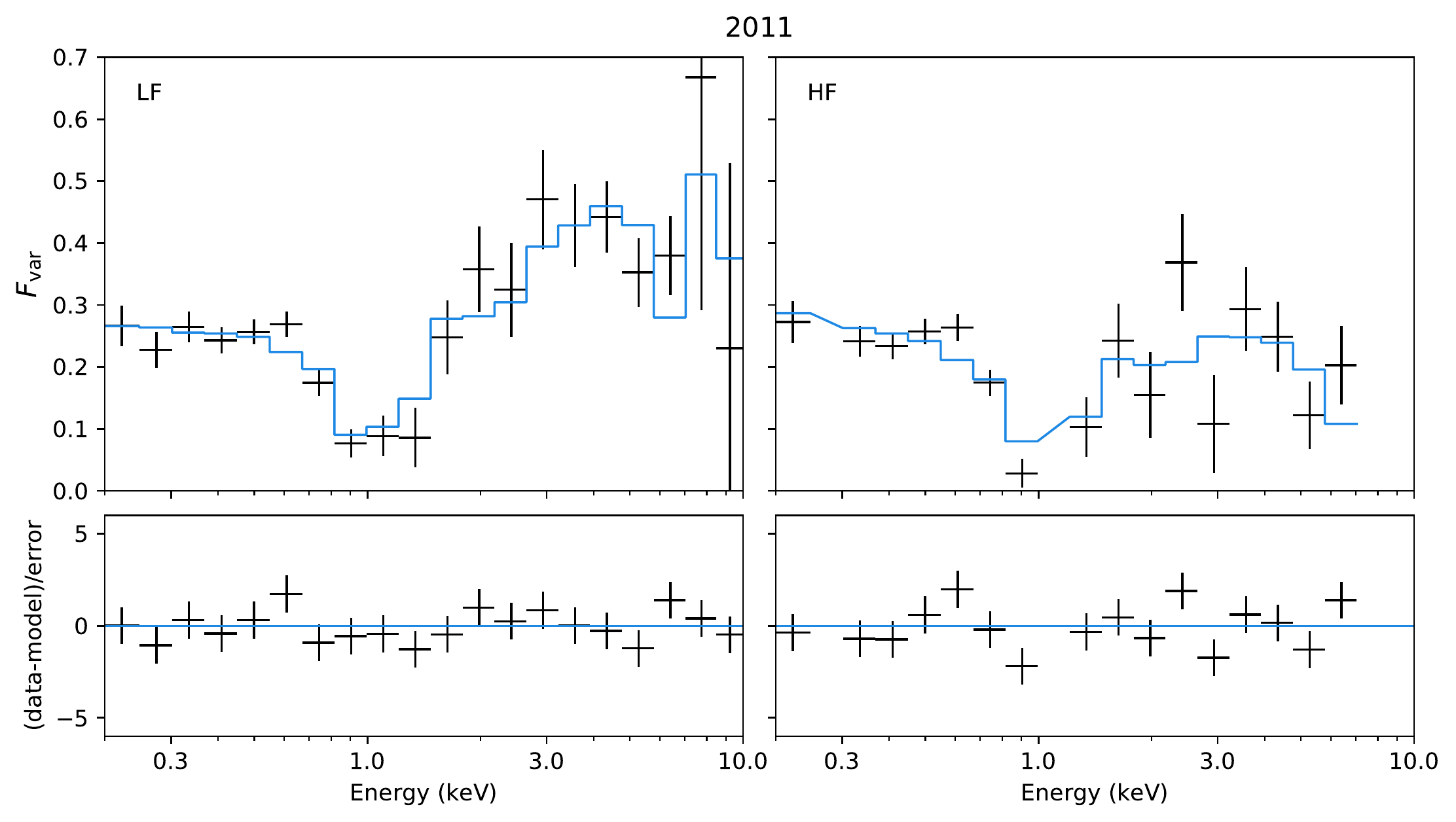}
    \caption{\fvar\ spectra for the low flux epochs. In each case the high and low frequency fits are shown in the left and right panels, respectively. Top panels show the data and model (blue stepped line). Bottom panels show the residuals in units of $\Delta\chi$. Both spectra show a dip in variance around 1~keV which is well modeled by a constant photoionised emission component. The noisier spectrum from 2000 in a similar low flux state is shown in Fig.~\ref{fig:lowflux_spectra2}}
    \label{fig:lowflux_spectra}
\end{figure*}

These lower flux spectra (shown in Fig.~\ref{fig:lowflux_spectra} and Fig.~\ref{fig:lowflux_spectra2}) are noisier due to lower signal, but some features can be seen in the spectra. Most notably, a strong dip in the variance is visible at 1~keV in the 2007b and 2011 spectra, and a weak dip is present in the same place in the 2000 spectrum. In general, negative features in $F_\mathrm{var}$ spectra requite the presence of a constant or less variable emission component. Constant multiplicative components (such as Galactic absorption) affect the absolute variability and flux equally, so have no effect on fractional variance. Constant emission, on the other hand, raises the flux while leaving the absolute variance unchanged, so $F-\mathrm{var}$ drops. In principle ionised absorption could produce a negative feature, but this would have to be fine-tuned and has not previously been observed in AGN. To produce a negative feature in the spectrum, the absorption would have to vary in such a way as to cancel out variability in the continuum. The absorption lines would have to get stronger as the flux rises, but not so much stronger that they start to introduce more variability. This would need to be quite a precise trend that is only present at low fluxes, and which is inconsistent with the observed trend of stronger UFOs at higher fluxes seen in 1H~0707-495 (Xu et al., in prep) and in other AGN \citep[][]{Parker17_nature,Parker18_pds456,Igo20}.

In this case, the 1~keV dip appearing in low flux spectra is naturally explained by the presence of constant photoionised emission, as discovered by \citet[][]{Kosec18}.
At this energy, the \textsc{fvar\_pidamp} model predicts a dip in variance for $\log(\xi)=2$--3 due to the presence of a large number of Ne, Mg and Fe emission lines (see section~\ref{sec:pion}). These lines raise the flux but not the variance, causing a drop in the fractional variance. A similar drop should be present at $\sim0.7$~keV, due to the O\textsc{viii} line, but this is not resolvable at CCD resolution. This interpretation also naturally explains why the dip feature is only present at low fluxes, as at high fluxes the continuum dominates and the contribution from photoionised gas is negligible.

An additional weak dip is visible at 6--7 keV in the 2011 model, but not in the data. This is caused by a predicted Fe~K$\alpha$ line in the model at high ionisations. The ionisation of the fit is driven by the low energy data, where the signal is higher, so the feature is not strongly disfavoured. We note that the \textsc{fvar\_pion} assumes that the continuum is a $\Gamma=2$ powerlaw, while the 2011 count spectrum shows a very strong relativistic iron line \citep{Fabian12}. This feature will lower the contribution of any photoionised emission in the high energy band, suppressing any corresponding negative feature in the data.

Aside from the 1~keV dip, the three spectra are relatively flat, with variance not concentrated either at soft or hard energies (unlike the 2019 spectrum). This suggests that in these epochs the short-term (within an observation) variability is at least partly intrinsic. Our intrinsic variability model, including the damping from the photoionised emission, describes the data well. However, the spectra are noisy, so we cannot exclude (or constrain) some contribution from absorption variance at low energies (see further discussion in Section~\ref{sec:absorption}). In particular, the 2000 spectrum (shown in Fig.~\ref{fig:lowflux_spectra2}) does not appear to show a drop in variance at low energies due to the soft excess, and it is possible that damping of the intrinsic variability at low energies is cancelled out by soft variability caused by absorption in the same energy band (see further discussion on this with a higher signal spectrum in Section~\ref{sec:longterm}).

\subsubsection{Soft variance: 2019}
\label{sec:2019}

\begin{figure*}
    \centering
    \includegraphics[width=0.7\linewidth]{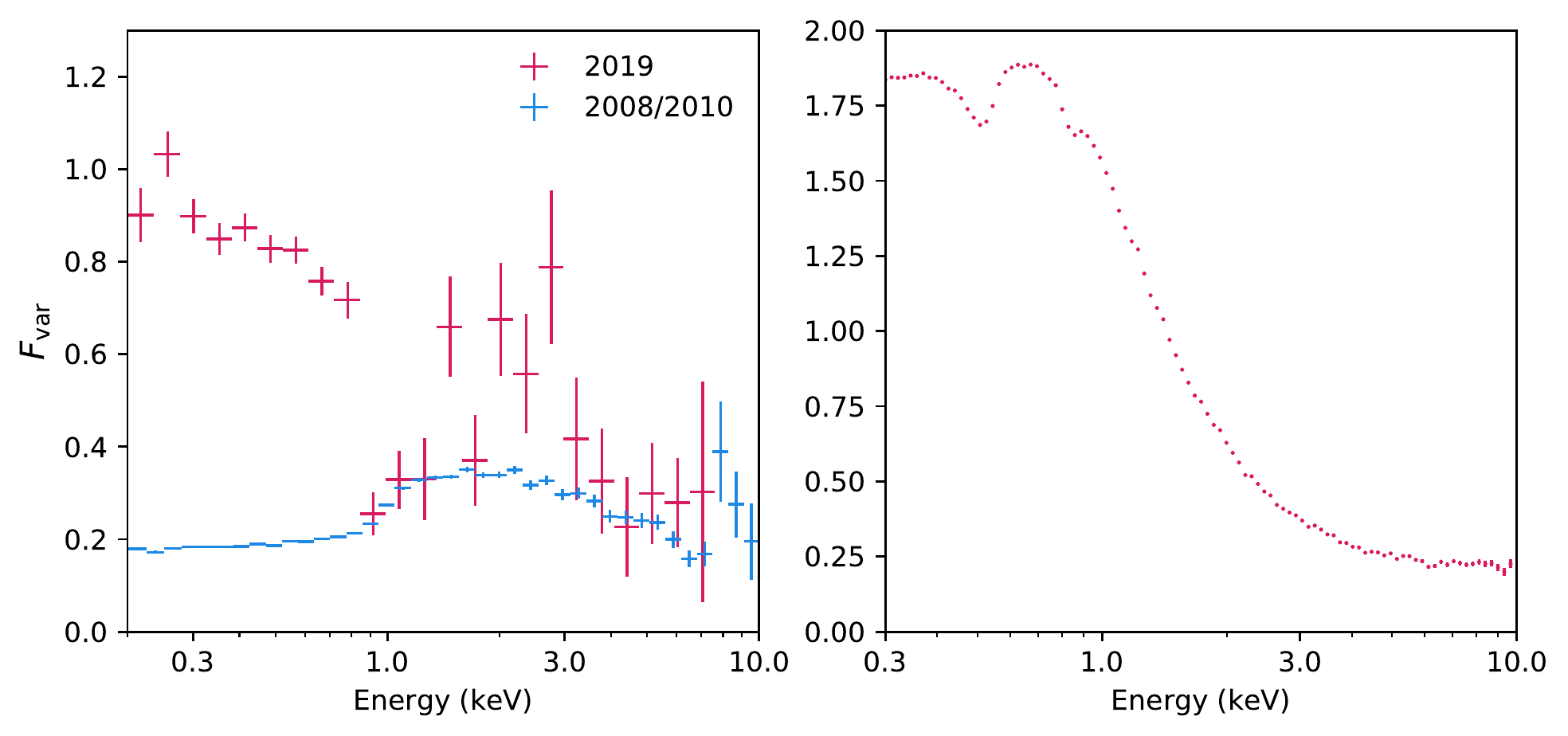}
    \caption{Left: Comparison of the 2019 low frequency variance spectrum (red) with the high flux 2008/2010 low frequency variance spectrum (blue) from Fig.~\ref{fig:highflux}. The 2019 spectrum is much more variable at low energies, and comparable above 3~keV. Right: The long-term variance spectrum of NGC~5548, adapted from \citet{Igo20}. This spectrum is dominated by the variance caused by long-timescale obscuration events, which cause dramatic variability at low energies. Note that structure from line-like drops in variance is visible at low energies, caused by the emission lines that become visible as the primary continuum is obscured.}
    \label{fig:2019comparison}
\end{figure*}

The 2019 spectrum observed by \erosita\ \citep[see Fig. 3 of][]{Boller21} is drastically different from the other spectra, as it is much more variable at low energies. The simultaneous \xmm\ spectrum shows the same basic shape. It has the same strong dip at 1~keV seen in the low flux spectra (2000, 2007b, and 2011), but the underlying variance is much higher at low energies. For comparison, we plot the 2019 spectrum and the 2008/2010 bright state spectrum on the same axes in Fig.~\ref{fig:2019comparison}, and a second plot of a source that is well known to be dominated by obscuration variability, which shows a strikingly similar qualitative spectrum \citep[NGC5548, see e.g.][]{Kaastra14}. It is also the lowest flux of all the \xmm\ observations: $1.6\times10^{-13}$~erg~cm$^{-2}$~s$^{-1}$ and $2.4\times10^{-13}$~erg~cm$^{-2}$~s$^{-1}$ for the 0.5--2 and 2--10~keV bands, respectively.

The intrinsic variability model does give an acceptable fit to the spectrum (Fig.~\ref{fig:spectra_2019_2}, top panel, Table~\ref{tab:2019_intrinsic}). However, the fit parameters are very different from those found for the other spectra. The powerlaw standard deviation is much higher, as is the index correlation. This is likely required to produce the very high variance at low energies. To compensate for this at high energies, the reflected fraction is high and the flux correlation very low, leading to a high degree of damping from the almost constant reflection. While the model can accommodate this spectrum, we consider it unlikely that this is the correct interpretation of the data.

\begin{figure*}
    \centering
    \includegraphics[width=0.7\linewidth]{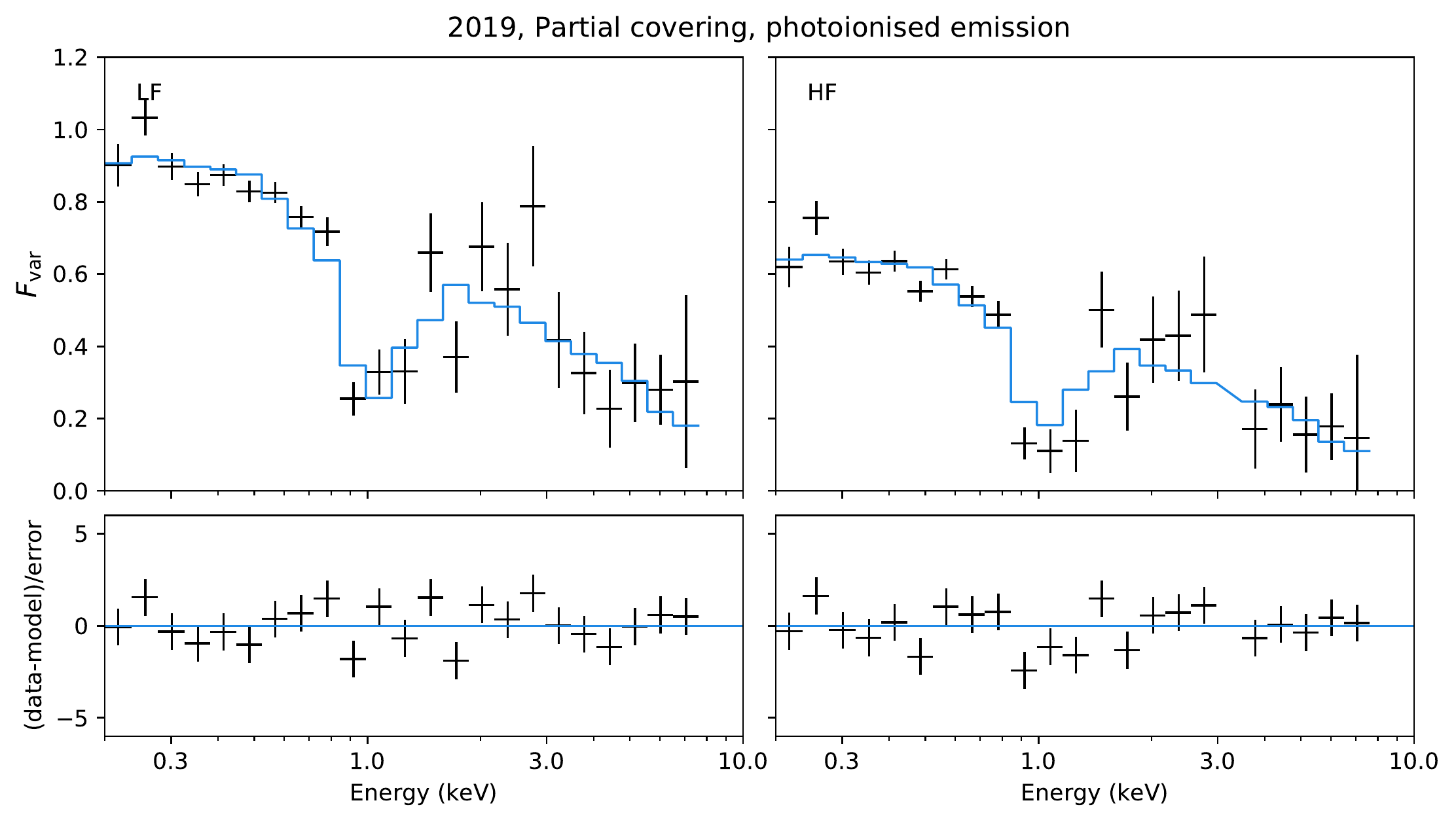}
    \caption{Photoionisation damped absorption fit to the 2019 \fvar\ spectra, fit with the model \textsc{fvar\_pion$\times$fvar\_pcov}. This fit gives a good description of the data, and the absorption interpretation explains the qualitative differences in the shape of the variance spectrum between the 2019 observation and the other epochs. Additional plots showing the fits with the intrinsic variability model and absorption without photoionised emission are presented in Fig.~\ref{fig:spectra_2019_2}}
    \label{fig:spectra_2019}
\end{figure*}

\begin{table}
    \centering
    \caption{2019 intrinsic variability fit parameters.}
    \begin{tabular}{l c c r}
    \hline
    \hline
    Parameter & LF & HF & Description\\
    \hline
$f_\mathrm{PI}$ & $ 2.9^{+0.2}_{-0.3} $ & & Photoionised fraction\\
$\log(\xi_\mathrm{PI})$ & $ 2.9 \pm 0.05 $ & & Ionization (erg~cm~s$^{-1}$)\\
$f_\mathrm{Ref}$ & $ 0.7^{+0.2}_{-0.1} $  & & Reflected fraction \\
$c_\mathrm{Ref}$ & $ 0.1 \pm 0.1 $ & & Flux correlation\\
$\log(\xi_\mathrm{Ref})$ & $ 1.51^{+0.1}_{-0.08} $ & & Ionization (erg~cm~s$^{-1}$) \\
$f_\mathrm{BB}$ & $ 0.9 \pm 0.1 $ & & Black body fraction\\
$\sigma$ & $ 0.78^{+0.04}_{-0.03} $ & $ 0.68^{+0.06}_{-0.1} $ & Powerlaw flux $\sigma$\\
$c_\Gamma$ & $ 1.27^{+0.05}_{-0.1} $ & $ 1.56^{+0.09}_{-0.05} $ & Index correlation\\
    $\chi^2$/dof & 59/35 &  & Fit statistic\\

    \hline
    \hline
    \end{tabular}
    \label{tab:2019_intrinsic}
\end{table}

We next consider a partial covering model, without accounting for the constant photoionized emission. This gives a poor fit to the data ($\chi^2$/dof of 220/38), fitting well at low energies but leaving large residuals at the 1~keV drop and at high energies where not enough variance is predicted (Fig.~\ref{fig:spectra_2019_2}, bottom panels). Adding in the damping from constant photoionised emission improves the fit hugely ($\chi^2$/dof of 65/36, Fig.~\ref{fig:spectra_2019}), and removes the strong residuals around the 1~keV feature. This model is consistent with a scenario where the primary X-ray emission from the accretion disk/corona is obscured, but the photoionised emission comes from a more extended region and is not significantly covered by the obscuring cloud.

\begin{table}

    \caption{Best fit parameters for the 2019 variance spectrum using the partial covering absorption model. Parameters marked with '*' are unconstrained.}
    \label{tab:fits_2019_abs}
    \centering
    \begin{tabular}{l c c r}
    \hline
    \hline
    Par. & LF & HF & Description\\
    \hline
\multicolumn{4}{|c|}{1. Partial covering}\\   
$N_\mathrm{H}$  & $ 2.0 \pm 0.2 $ && Column density ($10^{22}$~cm$^{-2}$)\\
$f_\mathrm{cov}$ & $ 0.81 \pm 0.02 $ & & Covering fraction\\
$\sigma_{f_\mathrm{cov}}$ &$ 0.3 \pm 0.06 $ & $ 0.18 \pm 0.009 $ & Covering fraction $\sigma$\\
$c_{N_\mathrm{H}}$ & $ >0.84 $ & $ >0.1 $ & Column density correlation\\
$\log{\xi_\mathrm{Abs}}$ & $ <-0.92 $ & & Ionization (erg~cm~s$^{-1}$)\\
$\chi^2$/dof & 483/38 & & Fit statistic\\
\hline

\multicolumn{4}{|c|}{2. Partial covering and photoionised emission}\\
$f_\mathrm{PI}$&  $  1.0 \pm 0.1 $ & & Photoionised fraction\\
$\log(\xi_\mathrm{PI})$ & $ 2.96^{+0.04}_{-0.07} $ & & Ionization (erg~cm~s$^{-1}$)\\
$N_\mathrm{H}$ & $ 15^{+3}_{-2} $  & & Column density ($10^{22}$~cm$^{-2}$)\\
$f_\mathrm{cov}$ & $ 0.74^{+0.05}_{-0.06} $ & &  Covering fraction \\
$\sigma_{f_\mathrm{cov}}$ & $ >0.43 $ & $ 0.24^{+0.07}_{-0.1} $ & Covering fraction $\sigma$\\
$c_{N_\mathrm{H}}$ & * & * & Column density correlation \\
$\log{\xi_\mathrm{Abs}}$ & $ <0.92$ & & Ionization (erg~cm~s$^{-1}$)\\
$\chi^2$/dof & 65/36 & & Fit statistic\\

    \hline
    
\multicolumn{4}{|c|}{3. Partial covering, photoionised emission, and powerlaw}\\
$f_\mathrm{PI}$ & $ 1.0^{+0.2}_{-0.1} $ & & Photoionised fraction \\
$\log(\xi_\mathrm{PI})$ &$ 2.81^{+0.07}_{-0.08} $ & & Ionization (erg~cm~s$^{-1}$) \\
$N_\mathrm{H}$ & $ 3^{+1}_{-2} $ & & Column density ($10^{22}$~cm$^{-2}$)\\
$f_\mathrm{cov}$ & $ 0.8 \pm 0.1 $ & &  Covering fraction  \\
$\sigma_{f_\mathrm{cov}}$ & $ >0.51 $ & $ 0.27^{+0.1}_{-0.02} $ & Covering fraction $\sigma$ \\
$c_{N_\mathrm{H}}$ &  * & *  & Column density correlation \\
$\log{\xi_\mathrm{Abs}}^1$ & $ 1.0^{+0.1}_{-0.2} $ & & Ionization (erg~cm~s$^{-1}$) \\
$\sigma_F$ & $ 0.12^{+0.02}_{-0.03} $ & $ 0.06 \pm 0.03 $ & Powerlaw flux $\sigma$\\
$\chi^2$/dof & 60/35 & & Fit statistic\\

\hline
\hline

    \end{tabular}
    
$^1$The error on the absorption ionization parameter returned from the standard xspec algorithm is underestimated here, due to a maximum in the $\chi^2$ contour just below $\log(\xi)=1$. When we explore the parameter space further using the \textsc{steppar} command, this value is instead an upper limit of $\log(\xi)<1.2$.
\end{table}

\begin{figure*}
    \centering
    \includegraphics[width=0.7\linewidth]{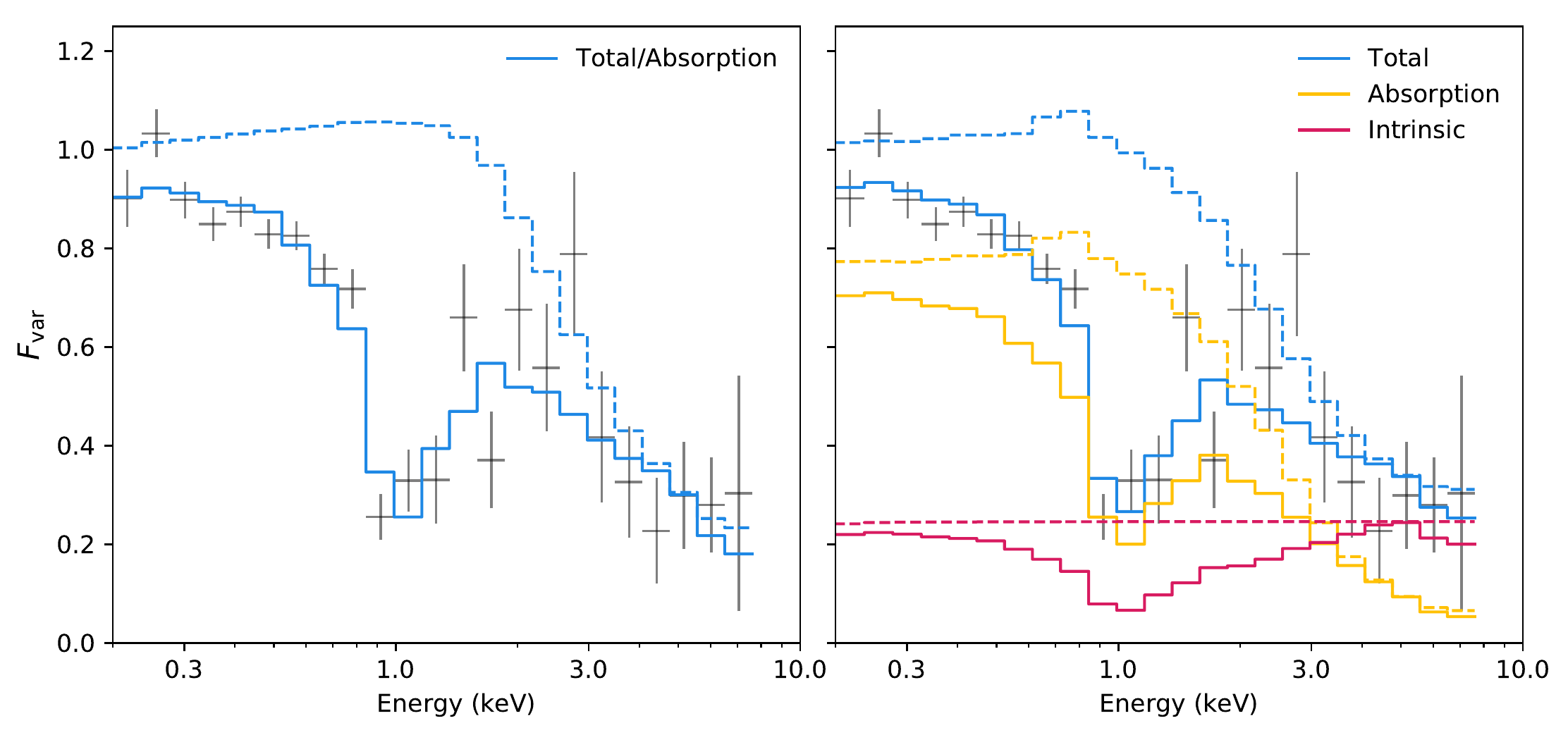}
    \caption{Model components for the partial covering and photoionised emission model (left) and the hybrid partial covering, powerlaw and intrinsic emission (assumed to be a powerlaw) model (right), fit to the low-frequency 2019 spectrum. In each case, dashed lines show the spectral shapes of the corresponding components when the damping effect of the constant photoionised emission is removed. In both cases, the photoionised emission is required in order to achieve a good description of the data. The statistical difference between the two is not highly significant, but the hybrid model is likely a more realistic scenario as some contribution from intrinsic variability at the same level as other observations is to be expected.}
    \label{fig:2019_breakdown}
\end{figure*}

While this absorption variance model provides an acceptable fit to the data, and leaves no obvious residual features, we note that even during an obscuration event the AGN likely shows some intrinsic variability as well. We therefore consider a hybrid model, with some contribution from intrinsic variability. Rather than including the full intrinsic variability model used above, we only add a powerlaw component (\textsc{Fvar\_pow}), with the index correlation frozen at 0 (i.e. with no pivoting). This allows the model to include some extra variance at high energies, where the absorption is less dominant, but only introduces one additional free parameter so the fit stays reasonably constrained. This does not improve the fit by a large amount ($\Delta\chi^2=5$, for two degrees of freedom), but is likely a better representation of the processes occurring than the pure absorption model as the intrinsic variability is unlikely to stop when the source is absorbed. A breakdown of the components of this model, and the model without an intrinsic component, is shown in Fig.~\ref{fig:2019_breakdown}. The absorption variability dominates below $\sim2.5$~keV, while at high energies the intrinsic variability is stronger. We note that in this model the level of the intrinsic variability ($F_\mathrm{var}\sim0.2$ in the low frequency spectrum) is consistent with that seen in the high flux spectra (Figs~\ref{fig:highflux},\ref{fig:highflux_spectra2}), which are well described by the intrinsic variability model. The best fit parameters for all absorption dominated models are given in Table~\ref{tab:fits_2019_abs}. 

\subsection{Long-timescale variance}
\label{sec:longterm}

The 2019 variance spectrum is qualitatively different from those in other epochs. We attribute this to the obscuration variability discovered by \citet{Boller21}, which seems only to contribute significantly in this observation. However, as \citeauthor{Boller21} point out, there are other low state count spectra of 1H~0707-495 that appear very similar to the 2019 spectrum, notably the 2011 spectrum. It is important to note that absorption will only leave signatures in the variance spectrum of an observation if it is variable within that observation. Therefore, there could be absorption present in the count spectrum while contributing nothing to the variance spectrum. 

To examine variability on longer (inter-observation) timescales, we next model the very low frequency spectrum, including data from all epochs (see details in data reduction). If there is variability in the absorption column on long timescales, it should increase the variance of this spectrum.

\begin{figure}
    \centering
    \includegraphics[width=\linewidth]{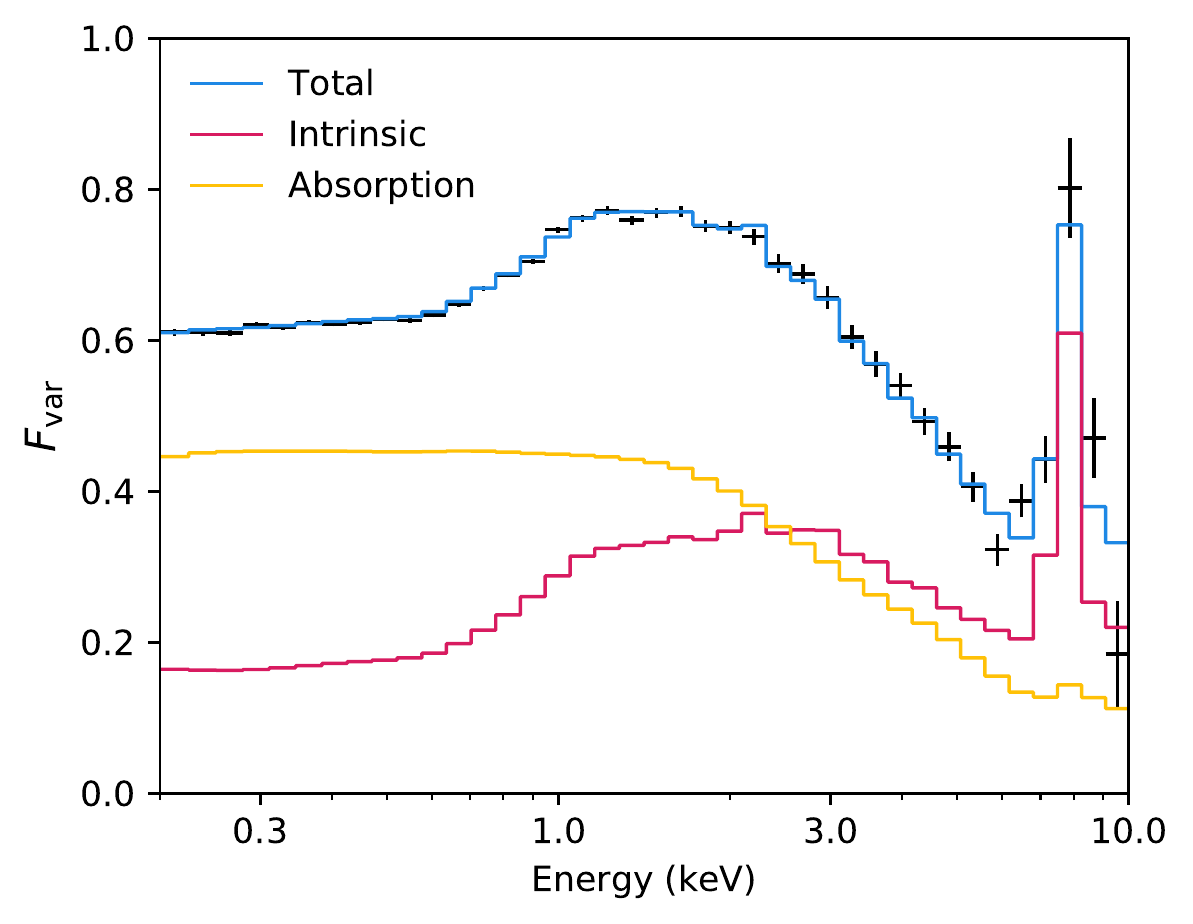}
    \caption{Very low frequency $F_\mathrm{var}$ spectrum. The spectrum is fit with a hybrid model containing both absorption variance and intrinsic variance. The low ionization absorption contributes most of the variance below 2~keV, while the intrinsic variability dominates at high energies, and gives the characteristic soft excess dip and UFO spike.}
    \label{fig:fit_lowfreq}
\end{figure}

The long-term variance spectrum is shown in Fig.~\ref{fig:fit_lowfreq}. It has some of the signatures of intrinsic variability: at low energies, the variability is damped by the less variable soft excess, and at high energies the Fe UFO line is clearly present, likely indicating the same kind of intrinsic continuum driven variability seen in other spectra and other sources. However, the overall spectral shape is significantly softer than the high frequency spectrum from the 2008 and 2010 campaigns (Fig.~\ref{fig:highflux}), suggesting that another process contributes additional variance at low energies. The two most obvious candidates for this are neutral absorption, as in the 2019 observation, or additional pivoting of the powerlaw at very low frequencies leading to more soft variance.

We fit this spectrum with a hybrid variance model, with both partial covering absorption and intrinsic powerlaw driven variability (\textsc{gsmooth $\times$ (fvar\_ufo $\times$ fvar\_bbdamp $\times$ fvar\_refdamp $\times$ fvar\_pow + fvar\_pcov)}). For simplicity, and because there is no 1~keV dip in this spectrum, we exclude the photoionisation damping component for this fit. We also fix the density and ionisation parameters of the reflection model to their best-fit values from the fit to the high flux/high resolution spectrum. The best-fit parameters are given in Table~\ref{tab:lowfreq_bestfit}.

\begin{table}
    \centering
    \caption{Best-fit parameters for the low frequency spectrum shown in Fig.~\ref{fig:fit_lowfreq}.}
    \begin{tabular}{l c c r}
    \hline
    \hline
    Parameter & Value & Description\\
    \hline
    
    $\log(\xi_\mathrm{UFO})$ & $ 4.6^{+0.08}_{-0.03} $ & Ionization (erg~cm~s$^{-1}$)\\
    $N_\mathrm{H}$ & $ >8.1 $ & Column density ($10^{24}$cm$^{-2}$)\\
    $c_\mathrm{UFO}$ &  $ 0.2^{+0.009}_{-0.04} $ & Ionization correlation\\
    
    $f_\mathrm{Ref}$  & $ 0.36^{+0.06}_{-0.02} $ & Reflected fraction\\
    $c_\mathrm{Ref}$ & $ <0.15 $ & Flux correlation\\
    
    kT & $ 0.120^{+0.004}_{-0.003} $  & Temperature (keV) \\
    $f_\mathrm{BB}$ & $ 0.31^{+0.09}_{-0.03} $  & Black body fraction\\
    
    $\sigma$ & $ 0.223^{+0.02}_{-0.01} $ & Powerlaw flux $\sigma$ \\
    $c_\Gamma$ & $ 0.36^{+0.09}_{-0.1} $ & Index correlation\\
    
    $N_\mathrm{H}$ & $ 6^{+1}_{-2} $  & Column density ($10^{22}$~cm$^{-2}$)\\
    $f_\mathrm{cov}$ & $ 0.1 \pm 0.1 $ &  Covering fraction  \\
    $\sigma_{f_\mathrm{cov}}$ & $ 0.62^{+0.09}_{-0.1} $ & Covering fraction $\sigma$ \\
    $c_{N_\mathrm{H}}$ &  $ >0.67 $  & Column density correlation \\
    $\log{\xi_\mathrm{Abs}}^1$ & $ 0.99^{+0.07}_{-0.3} $ &  Ionization (erg~cm~s$^{-1}$) \\
    $\chi^2$/dof & 64/38 & Fit statistic\\
    
    \hline
    \hline
    \end{tabular}
    \label{tab:lowfreq_bestfit}

$^1$The error on the absorption ionization parameter returned from the standard xspec algorithm is underestimated here, due to a maximum in the $\chi^2$ contour just below $\log(\xi)=1$. When we explore the parameter space further using the \textsc{steppar} command, this value is instead an upper limit of $\log(\xi)<1.1$.

\end{table}

This hybrid model gives a good description of the data, with the variance below $\sim2$~keV dominated by absorption and above by intrinsic variance. Both components appear to be essential. Removing the absorption variance worsens the fit by $\Delta\chi^2=22$, for 5 degrees of freedom, and leaves large residuals in the Fe band as the intrinsic variance model cannot simultaneously fit both the low energy and high energy bands. Removing the intrinsic variance component (including the UFO) worsens the fit drastically, by $\Delta\chi^2=348$, for 11 degrees of freedom. If we apply the UFO variance enhancement to the absorption (\textsc{fvar\_ufo $\times$ fvar\_pcov})\footnote{Note that the assumptions behind the UFO variance model (ionization response to a variable powerlaw continuum) are not valid with this setup, so it should not generally be used in this way, and this should be regarded as a phenomenological test.}, we still find a significantly worse fit ($\Delta\chi^2=72$, for 7 degrees of freedom), as the model cannot adequately fit the soft excess. 

To test that the requirement for both intrinsic and absorption variance is not driven purely by the soft variance of the 2019 data, we recalculated the spectrum with the 2019 data excluded. This did not lead to any significant changes in spectral shape, so we conclude that the long term variance includes this absorption component independently of the 2019 observation.

\section{Discussion}

We have fit the variance spectra of 1H~0707-495 with various physically-motivated models, aiming to understand the nature of the extreme variability seen in this source. We have fit each epoch separately, and the resulting spectra can be broadly divided into three categories:
\begin{enumerate}
    \item High flux spectra, dominated by intrinsic variability and enhanced by the UFO (2002, 2007a, 2007c, 2008, 2010). These spectra peak in variance around 1.5~keV, where the powerlaw contribution is greatest, and show a strong dip in variance at low energies due to the soft excess. Those epochs with the highest signal also show peaks in the variance corresponding to the UFO lines, as seen in IRAS~13224-3809 and PDS~456. None of these spectra show the 1~keV dip seen in the low flux spectra, as the contribution from photoionised emission is negligible.
    \item Low flux spectra, where the intrinsic variance is damped by the constant photoionised emission (2000, 2007b, 2011). These spectra are consistent with the high flux spectra, but have a much higher contribution from the photoionised emission lines, causing a drop in variance around 1~keV. 
    \item The 2019 low flux spectrum. This is qualitatively different, showing drastically more variance at low energies than the other epochs. This is likely due to obscuration, which enhances the variance at low energies, modified by the photoionised lines, which cause a sharp drop in variance at 1~keV. 
\end{enumerate}

\subsection{The unusual variability in 2019}

\citet{Boller21} were able to fit flux-resolved \erosita\ and \xmm\ spectra from the joint 2019 observation with a model consisting of partial covering absorption and relativistic reflection, and attribute the differences between the three spectra primarily to absorption. We note, however, that the ionization parameter of the reflection in their model is also free to vary between the three spectra (along with the \textsc{relxill} normalization and the covering fraction of the absorber), and the effect of this is crucial to understanding how the model is able to fit the spectra.

\begin{figure}
    \centering
    \includegraphics[width=0.8\linewidth]{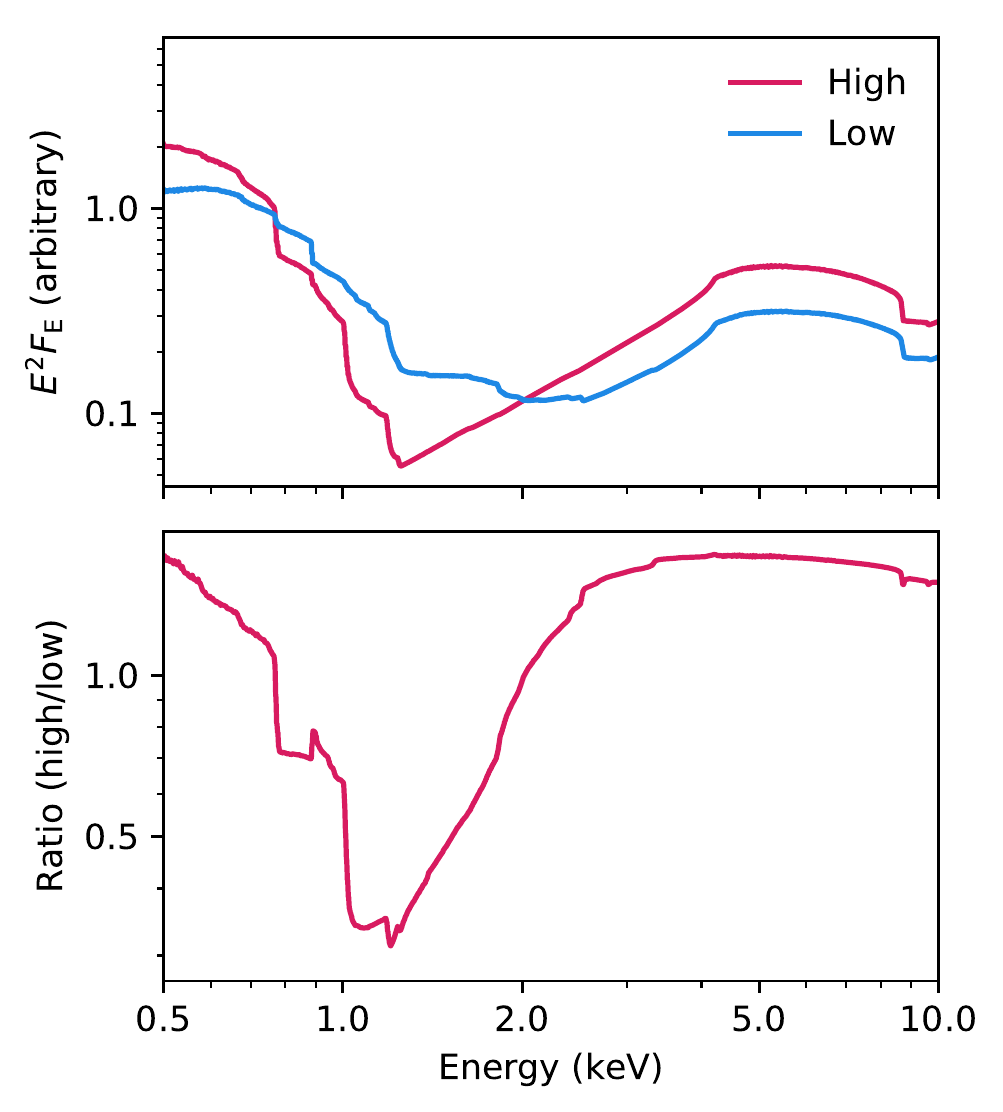}
    \caption{Top: Reflection spectra calculated with \textsc{relxill} \citep[][]{Garcia14} with the best-fit parameters from \citet{Boller21} for the low and high flux spectra. Bottom: ratio between the high and low flux spectra. The low flux spectrum is a factor of 3 brighter than the high flux spectrum at 1~keV.}
    \label{fig:boller_reflection}
\end{figure}

In their best fit model, the ionisation of the reflection is 0.68 and 0.64 in the high and medium flux spectra, respectively, but 1.74 in the low flux spectrum. This implies an increase of the disk surface ionisation by over an order of magnitude in the low state, coinciding with the increase in absorption column. Spectrally, the main effect of this increase is to drastically increase the flux of the reflection spectrum at 1~keV in the low flux spectrum (Fig.~\ref{fig:boller_reflection}).

This increase in the reflection flux means that there is much less variability at 1~keV than would be predicted in the absorption model, enabling it to fit the spectrum. \citet{Boller21} suggest that the change in reflection ionisation in their fits is likely not correct, and attribute it to un-modelled changes in the ionization of the absorber, which they assumed to be neutral. However, as we have demonstrated (see Fig.~\ref{fig:pcov_models}, bottom panel), increasing the ionization of the absorption decreases the amount of low-energy (<1~keV) variability, and increases the variance at 1 keV where strong lines are produced (hence higher ionizations are disfavoured in our modelling), so such a model would not be able to explain the variability.

It is possible that the ionization of the accretion disk could rise in low flux states in some scenarios. For example, in a lamp-post light-bending scenario \citep[e.g.][]{Miniutti03,Miniutti04, Parker14_mrk335}, the corona contracts close to the accretion disk and more of the primary emission is focused onto the smallest radii. This could lead to higher ionization being observed in this case. However, focussing the same amount of emission onto a smaller region is unlikely to lead to such a dramatic increase in ionisation, and the increased radiation pressure may act to increase the surface density and lower the ionisation (Fabian et al., in prep), so this solution is also unlikely.

A hybrid model, where the disk ionisation coincidentally increases during an absorption-induced low state could explain a single observation of this phenomenon, however the 1~keV dip is observed in every low state, regardless of whether the variance is dominated by absorption or intrinsic variability. This is trivially explained by constant photoionized emission, which produces strong lines at 1~keV and will only have an effect on the variance spectrum when the source flux is low.

This does not imply that the absorption interpretation for the 2019 data is incorrect. As we show in Fig.~\ref{fig:spectra_2019}, a partial-covering model can explain the variance spectrum with the addition of constant (i.e. unabsorbed) photoionised emission. Additionally, the underlying variance spectrum is very soft and very high amplitude compared to the other observations. None of the bright state spectra have variances at 0.5~keV above $\sim0.5$ because of heavy damping from the soft excess, while in 2019 it approaches 1. While this variability can be well fit with our intrinsic variability model, the parameters required differ strongly from those in other observations, and the signatures of intrinsic variability (soft excess damping, and the UFO lines) are not present.

One further thing to note when considering the 2019 spectrum is that the \xmm\ response is highly non-diagonal at low energies. As discussed in \citet{ingram19}, this means that a significant number of photons in the 0.3--0.7~keV band may instead be misclassified photons with energies of $\sim1$~keV, particularly with spectrally hard sources. In this case, this means that the variability below 1~keV may actually be even more extreme, as it will be slightly diluted by low variability 1~keV emission. This is likely a relatively mild effect, and will not qualitatively change our conclusions, but possibly introduces a slight bias into the absorption parameters. Since the constraints on these parameters are fairly weak in our fits it is not likely to be important here, but this effect should be considered when studying soft variability with \xmm\ in general.

\subsection{The role of absorption variability}
\label{sec:absorption}

We agree with the interpretation of \citet{Boller21} that the unusual variance seen in 2019 is likely due to absorption. This establishes a clear precedent for absorption induced variability in 1H~0707-495, and raises the question of what role absorption variability plays in other observations of the source.

The other individual epoch variance spectra are qualitatively different from the 2019 spectrum, with much lower soft variance and signatures of intrinsic variability, and are well described by the intrinsic variability model. In general, we cannot rule out some contribution from absorption variance within individual epochs, particularly those with low fluxes and poor signal, but with the exception of 2019 it cannot be the dominant factor.

On longer timescales (between observations) the picture is different. Our long term variance spectrum (Fig.~\ref{fig:fit_lowfreq}) shows significantly more soft variance than the individual epoch spectra. While it has the signatures of intrinsic variability, it suggests that absorption variability also contributes. Fitting this spectrum with a hybrid model suggests that the long-term variability below $\sim2$~keV is dominated by absorption variance, and above 2~keV by intrinsic variance (enhanced by the UFO).
This is consistent with the expected behaviour of low-ionisation absorption - it must be located far from the black hole, otherwise it would be highly ionised, and therefore it cannot produce much high-frequency variability. We note that the 2019 observation is the only observation that shows clear evidence for absorption variability on single-orbit timescales, and that the variability in this observation is dominated by a gradual decline in flux \citep[Fig. 1 and 2 of][]{Boller21}, rather than the spiky stochastic variability more commonly seen in 1H~0707-495.

Overall, our modelling suggests a scenario where the variability on short timescales is dominated by the intrinsic behaviour of the AGN, while on longer timescales obscuration contributes significantly. In 2019, we witness the onset of an obscuration event which dominates the variance while in other low states like that in 2011, which is spectrally very similar to 2019, the onset of obscuration occurs outside the observation and the absorption is approximately constant during the observation. Due to the different timescales involved, this is unlikely to be a problem for high frequency timing results \citep[such as the detection of Fe~K reverberation by][]{Kara13_1h0707}. This is consistent with the properties of the RMS-flux relation observed in 1H~0707-495, IRAS~13224-3809 and accreting compact objects in general. The RMS-flux relation occurs over a broad range of frequencies, meaning that variability on different frequencies must be linked. This occurs naturally in the multiplicative propagating fluctuations model \citep[][]{uttley05}, where low frequency variability modulates high frequencies, but not in an additive model such as when the variability arises from multiple discrete clouds of absorbing gas. In practice, this means that absorption variability cannot dominate in general, and cannot be the origin of the high amplitude, high frequency variability seen in sources like 1H~0707-495 \citep[see discussion in][]{Alston20_natast}. The RMS-flux relation in IRAS~13224-3809 will be examined in more detail in Alston et al. (in prep).

We also note that we find no evidence of the kind of extreme column density partial-covering absorption invoked by some authors as an alternative explanation for the Fe~K emission line in 1H~0707-495 and other sources \citep[e.g.][]{Mizumoto14}. These models typically require low ionization, near Compton-thick material to vary extremely rapidly to produce the observed high frequency variability, while never either fully covering or fully un-covering the source. We consider it unlikely that such material exists, and our absorption models strongly rule out such high column densities both in the 2019 fit and in the long-term variability.

\subsection{Photoionised emission}

The main difference between the intrinsic variance spectra of 1H~0707-495 and IRAS~13224-3809 is seen at 1~keV, where the low flux observations of 1H~0707-495 show a strong dip. This dip is present in both the observations that we interpret as dominated by intrinsic variability and the 2019 observation that is likely dominated by obscuration. This suggests an origin independent of the inner accretion flow and the obscurer.

Producing such a drop requires a constant or low variability emission component, as a constant absorption component leaves no signatures in the fractional excess variance. We suggest that this emission can be associated with the blueshifted photoionised gas identified by \citet{Kosec18}, which notably has no counterpart in IRAS~13224-3809 \citep[e.g.][]{Pinto18}. \citeauthor{Kosec18} performed a flux-resolved analysis, and found the emission in their lowest flux spectrum. This agrees well with our variance analysis, where the 1~keV variance drop is only present at low fluxes. The emission component detected by \citeauthor{Kosec18} has an ionization of $\log(\xi)=2.4$~erg~cm~s$^{-1}$, consistent with the $\log(\xi)=2$--3~erg~cm~s$^{-1}$ that we find from fitting the variability spectra with a photoionisation damping model, and the ionisation range that produces a strong emission feature at $\sim1$~keV. We note that high ionizations ($log(\xi)\sim3$) are ruled out by \citeauthor{Kosec18} as they predict a narrow Fe emission line that is not observed in the data. Some of our fits have ionizations in this regime, although they are not generally well constrained. We note that the ionization in our fits is likely somewhat degenerate with the velocity of the material, as the main effect of increasing ionization in the \textsc{fvar\_pion} is to shift the main dip feature to higher energies, as it becomes dominated by higher ionisation species. Since the velocity of this material is significant ($\sim8000$~km~s$^{-1}$), it is possible that shifts in the velocity between observations lead to us inferring different ionisations.

We note that we are sensitive to different features than the RGS data examined by \citeauthor{Kosec18}. The RGS spectrum is dominated by the O\textsc{viii} line at 0.65~keV, with only weak emission from the Ne/Fe complex that causes the dip we observe at 1~keV. There are several effects in play here. Firstly, the approximately constant energy resolution of CCDs mean that their resolving power declines rapidly at low energies, so the O\textsc{viii} line cannot generally be resolved. The Ne and Fe complex at 1~keV is broad, and higher energy, so much easier for the pn to pick up. Secondly, as a grating the RGS is more sensitive to individual strong lines, while the pn picks up integrated flux from multiple lines in a broader band. Finally, the absolute flux in a feature is less important for an $F_\mathrm{var}$ spectrum than the relative flux of a feature to the continuum. The O\textsc{viii} line may have a larger absolute flux, but the Ne/Fe complex contributes a larger fraction of the total flux at 1~keV, leading to a much stronger feature in the variance spectrum. We note that this is a common result of comparing EPIC and RGS spectra -- see for example Figs 5 and 8 of \citet{Parker19_mrk335}, where the O lines dominate the RGS spectrum but the PN spectrum shows a broad bump at 1 keV and no clear evidence for O emission.

\citeauthor{Kosec18} suggest that the emission comes from a relatively large scale wind, driven by the smaller scale UFO. This likely means that the emission lines would be located on large enough scales that they are relatively unaffected by the obscuration that absorbs the emission from the inner accretion flow and corona. The emission region does not necessarily have to be on larger scales than the absorption, since the absorption is likely made of small clumps (as it partially covers the corona) and those clumps would cover a roughly constant fraction of a more extended photoionised emission region, leading to low variability. This explains why the feature is still present in the 2019 observation - as the majority of that emission is not affected by the absorber, it remains approximately constant while the absorbed continuum varies, causing lower variance in the band where the photoionised lines are present.


\subsection{The Ultra-Fast Outflow}

We note in Section~\ref{sec:highflux} that the UFO variance in 1H~0707-495 is significantly stronger than that observed in IRAS~13224-3809 and PDS~456 \citep[][]{Parker20,Haerer21}. The reason for this is not clear, but the relatively low value of the ionization correlation $c_\mathrm{UFO}$ suggests that it is not being driven by the UFO gas responding to the continuum. It does still seem to be associated with the intrinsic variability rather than the absorption variance -- no clear UFO feature is present in the 2019 spectrum, which is dominated by the absorption variability, and a strong feature is present in the bright 2008 and 2010 datasets. We suggest that the most likely reason for the higher variability in the UFO is that an additional process (or processes), in addition to ionization and recombination of the gas, contributes to the variability of the UFO line features. 

Flux resolved spectroscopy (Xu et al., in prep.) reveals that 1H~0707-495 does show a similar trend to IRAS~13224-3809 and PDS~456 \citep{Parker18_pds456} of decreasing absorption strength with increasing source flux, consistent with the gas becoming increasingly ionised and no longer producing line absorption. This means that at least some of the variance we see in the UFO lines is likely due to ionisation. Xu et al. also find an anti-correlation between wind velocity and X-ray flux, the opposite of what is observed in PDS~456 \citep[][]{Matzeu17} and IRAS~13224-3809 \citep[][]{Pinto18}, and speculate that it may be caused by over-ionisation of the gas leading to a drop in line radiation pressure, or a shift in the launch radius of the wind to larger radii at higher fluxes. While we have not modelled this, it is possible that this anticorrelation could lead to a further enhancement of the variance.
There are various other processes which could increase the observed variance in the UFO, such as clumps in the wind crossing the line of sight, or a changing mass outflow rate, which may still cause lines to weaken with the X-ray flux (note that any process that caused the lines to strengthen with flux would cause dips in the variance, not spikes).

\subsection{Frequency behaviour}
We do not have the level of signal required to track detailed properties of the source with frequency, but some general properties can be observed. In particular, the correlation between powerlaw flux and photon index, $c_\Gamma$, is typically weaker in the high frequency spectra. This parameter controls the degree of pivoting in the powerlaw, with higher values increasing the variance at low energies, leading to a softer spectrum.

Where this parameter is constrained for both low and high frequency spectra, the low frequency spectrum always has a stronger correlation, implying a soft intrinsic continuum variance spectrum, with the notable exception of the 2011 spectrum. This trend is common in accreting compact objects, and is generally thought to be due to fluctuations propagating through the corona \citep[e.g.][]{Mahmoud18}.

The 2011 low frequency spectrum is unusually hard, and does not show the same trend with frequency as the other epochs. The reason for this is not clear. The spectral hardness can be explained by the high fraction of photoionised emission in this observation, which suppresses the soft variance, but this does not explain why the high frequency spectrum is softer. One possibility is that the reflection spectrum suppresses the variance in the high frequency spectrum. The 2011 count spectrum shows a high reflection fraction, consistent with a compact X-ray source close to the event horizon \citep[][]{Fabian12}. The high frequency range we use overlaps with the reverberation frequency observed in 1H~0707-495 \citep[][]{Fabian09,Kara13_1h0707}, and the reverberation delay will act to reduce the variability in the reflection spectrum, leading to increased damping in this frequency range. This effect would be most noticeable at high energies where the reflection contributes strongly, leading to a softer variance spectrum at high frequencies.

No clear trends are visible in the absorption models with frequency, as we only have the 2019 spectrum to work with, and the parameters are generally poorly constrained and/or highly degenerate.
However, it is interesting to note that the high flux observations show a stronger drop in average $F_\mathrm{var}$ between the low and high frequency spectra. This drop is a natural consequence of the red-noise power spectrum common to accreting compact objects \citep[e.g.][]{gonzalezmartin12}, which produces less variance at high frequencies. This weaker trend with frequency may indicate that some process that does not follow the same red-noise power spectrum contributes at low fluxes. One possible interpretation of this is that absorption variability does not necessarily have to follow a red noise power spectrum, so if some part of the variance in the low flux epochs is due to absorption a weaker drop in variance with frequency will be observed. A full examination of the power spectra of each epoch will be presented by Alston et al. (in prep), which will reveal this behaviour in much more detail.

\section{Conclusions}

We have presented an analysis of epoch-resolved $F_\mathrm{var}$ spectra of the NLS1 AGN 1H~0707-495 with \xmm . We have developed two new variance models for fitting these spectra, for modelling absorption variance and damping from constant photoionised emission. In general, our fits provide a satisfactory explanation of the complex variability behaviour of 1H~0707-495. Our main results are summarised below:
\begin{itemize}
    \item Most epochs can be well described by an intrinsic variability model, where the powerlaw continuum drives the variance, and is damped by reflection and enhanced by the UFO.
    \item In low flux epochs, such as in 2011, a dip in the variance around 1~keV appears. This can be explained by the presence of constant flux ionised emission lines, which make up a much higher fraction of the flux in low states and dramatically decrease the fractional variance in this band. We model this damping effect with a photoionised emission model, and find that the gas properties are consistent with those identified by \citet{Kosec18}.
    \item In 2019, the variance is much softer than in other epochs, as noted by \citet{Boller21}. This suggests that the variance is qualitatively different. We fit this variance spectrum with a partial covering absorption model, but also require damping from the photoionised emission at 1~keV to fully describe the soft variance.
    \item In low flux epochs, the variance drops less with frequency. This may indicate the contribution of an additional process in these observations (such as absorption variability) which does not follow the same red-noise power spectrum as the intrinsic variability which dominates the high flux epochs.
    \item We also construct a long-term variance spectrum, which includes variance between observations. This spectrum can be well described by a hybrid model, including both absorption and intrinsic variance. Based on this, we conclude that aside from the 2019 observation absorption variance is generally restricted to low frequency variability, but may be responsible for some of the extreme low flux states seen in 1H~0707-495.
\end{itemize}

While variance spectra have been used extensively for qualitative analysis of AGN spectra, they have not generally been examined in a detailed qualitative fashion. In this work we have demonstrated the potential for this kind of detailed analysis to shed light on the nature of AGN variability, even in complex scenarios where multiple processes contribute to the total variance. As variance spectra are easy to calculate, and the models we have developed here are simple, generally applicable, and freely available, we are hopeful that this approach will be used more broadly for the analysis of AGN variability, as well as for other compact objects.

\section*{Acknowledgements}
We thank the referee for their detailed and constructive feedback. WNA is supported by an ESA research fellowship. J.J. acknowledges support from the fellowship of China Postdoctoral Science Foundation (2021M691822), the Tsinghua Shuimu Scholar Program and the Tsinghua Astrophysics Outstanding Fellowship.

\section*{Data availability}
All data used in this work is available from the XMM-Newton science archive (\url{http://nxsa.esac.esa.int/}). The models used are available from \url{https://www.michaelparker.space/variance-models}. The code used to generate them can be provided on request to the authors.




\bibliographystyle{mnras}
\bibliography{1h0707_bibliography,var} 




\appendix

\section{Additional figures}

\begin{figure*}
    \centering
    \includegraphics[width=0.7\linewidth]{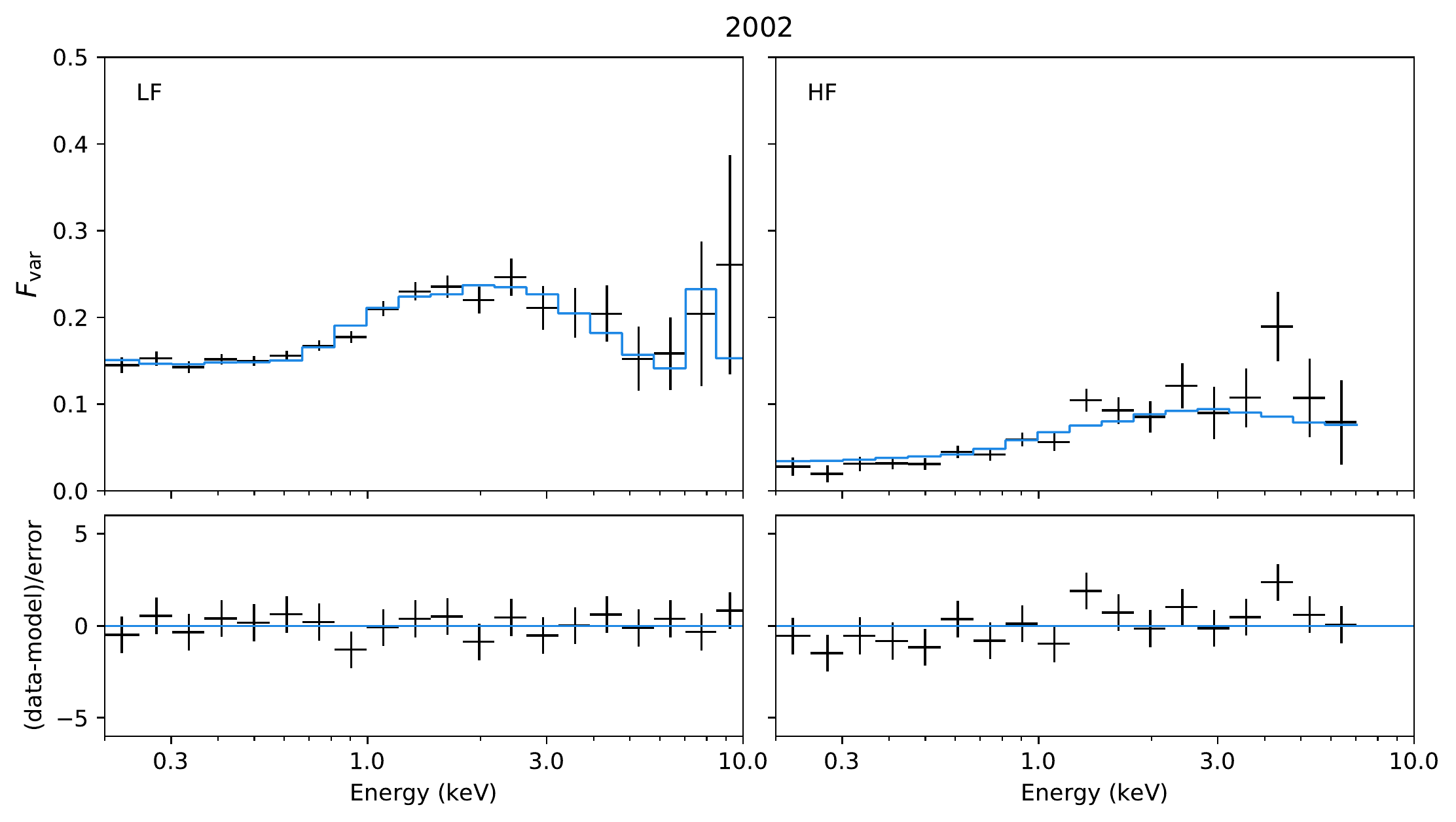}
    \includegraphics[width=0.7\linewidth]{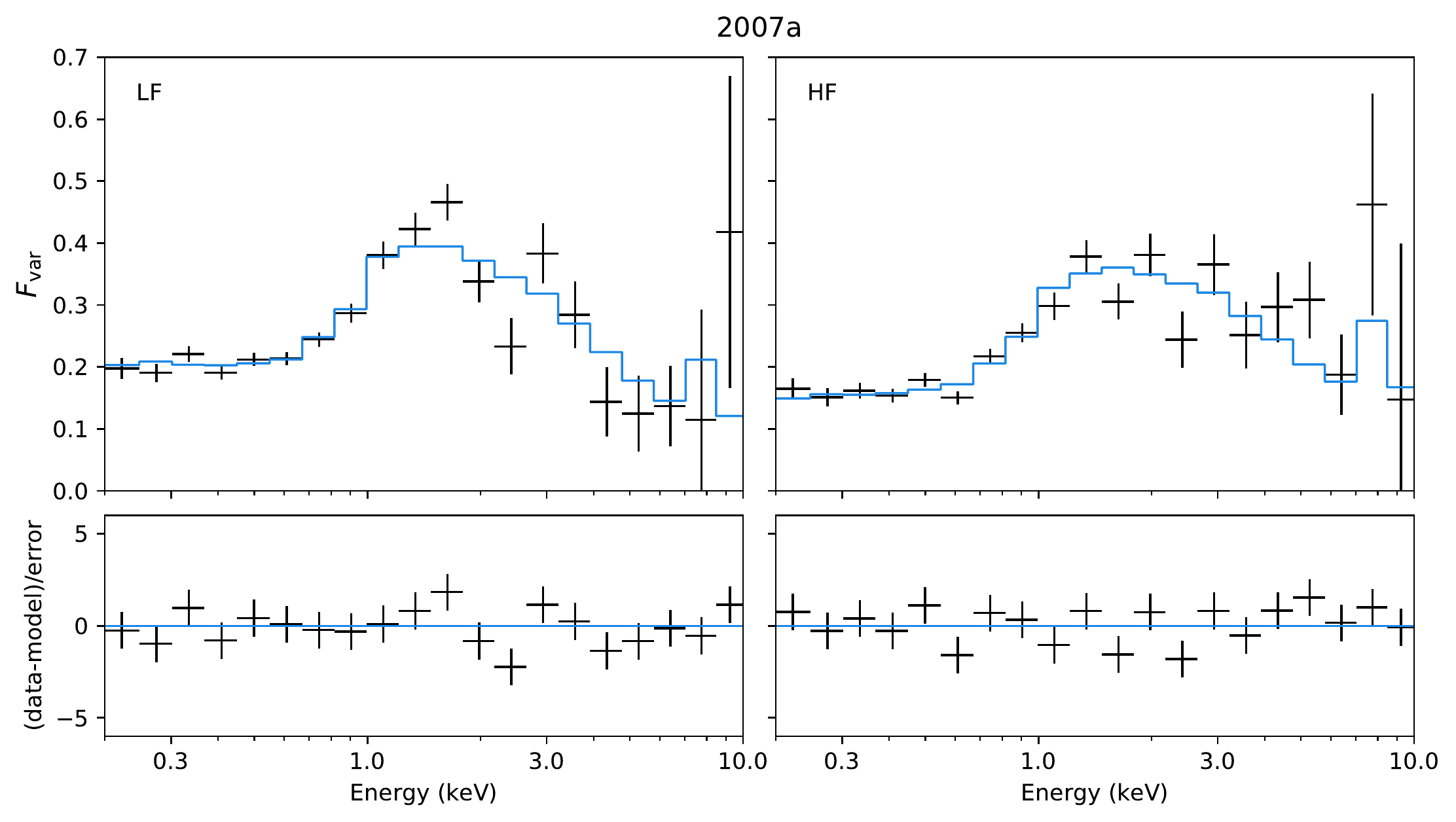}
    \includegraphics[width=0.7\linewidth]{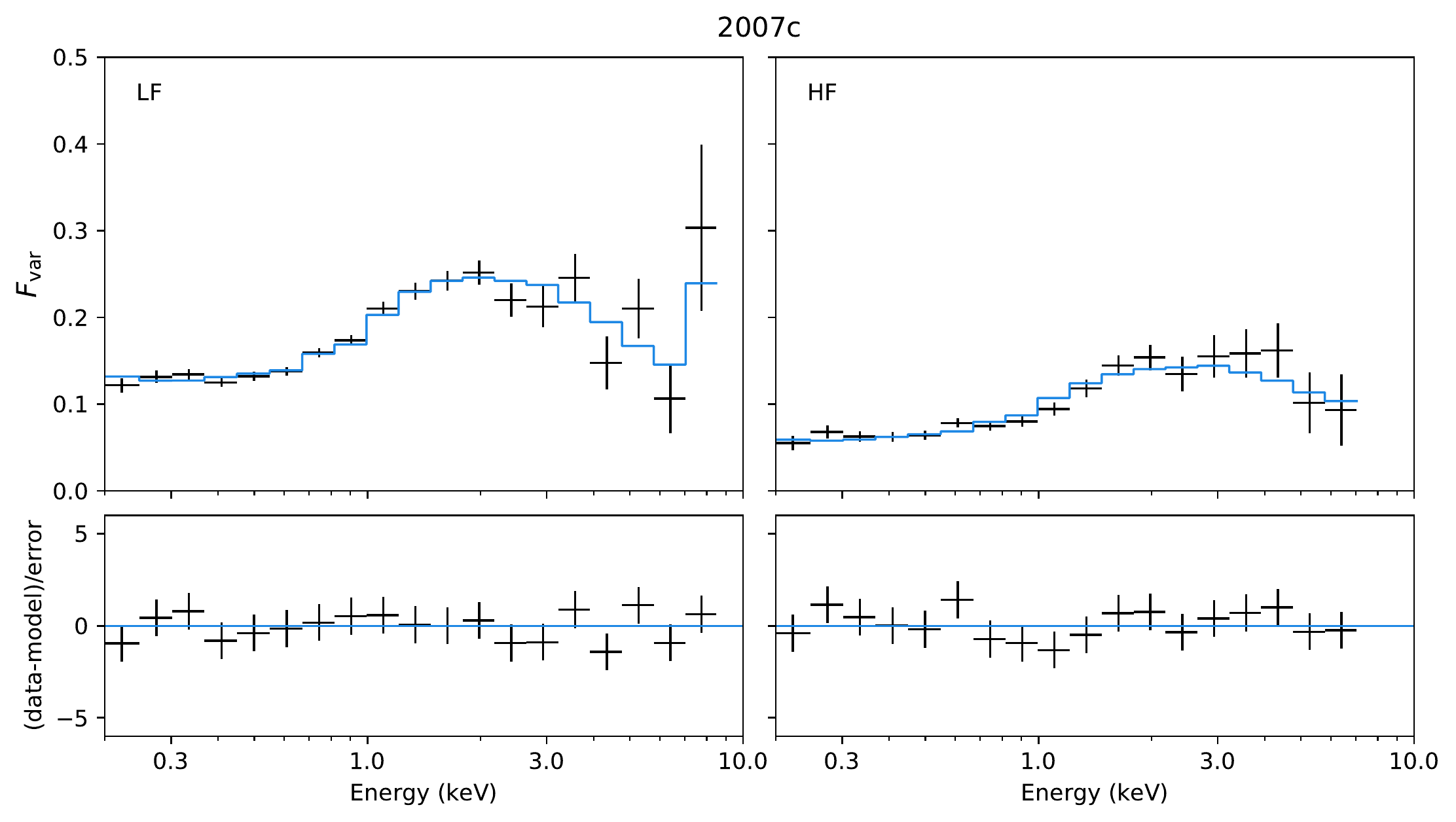}
    \caption{\fvar\ spectra for the lower signal (shorter exposure) high flux epochs. All three spectra show damping from the soft excess at low energies. They are consistent with the presence of UFO features, but the signal is limited by the short exposure times so the spikes are not significantly detected. There is some possible structure in the 2007a LF spectrum at $\sim1.5$ and $\sim3$~keV, but this is likely just noise.}
    \label{fig:highflux_spectra}
\end{figure*}

\begin{figure*}
    \centering
    \includegraphics[width=0.67\linewidth]{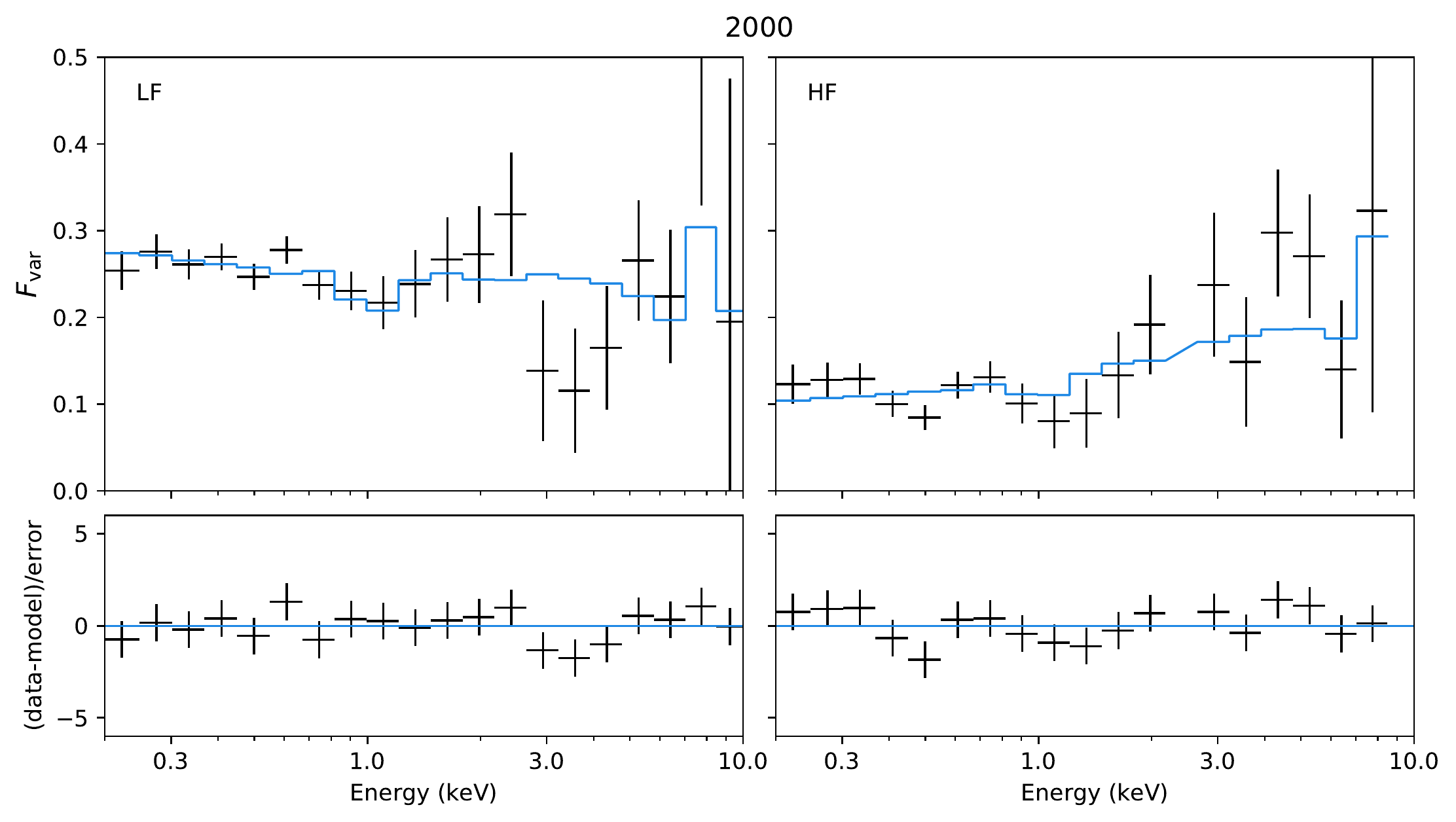}
    \caption{\fvar\ spectra for the low flux 2000 observation. Similar to the other low flux spectra, although weaker, this observation shows a dip in the variance around 1~keV.}
    \label{fig:lowflux_spectra2}
\end{figure*}

\begin{figure*}
    \centering
    \includegraphics[width=0.67\linewidth]{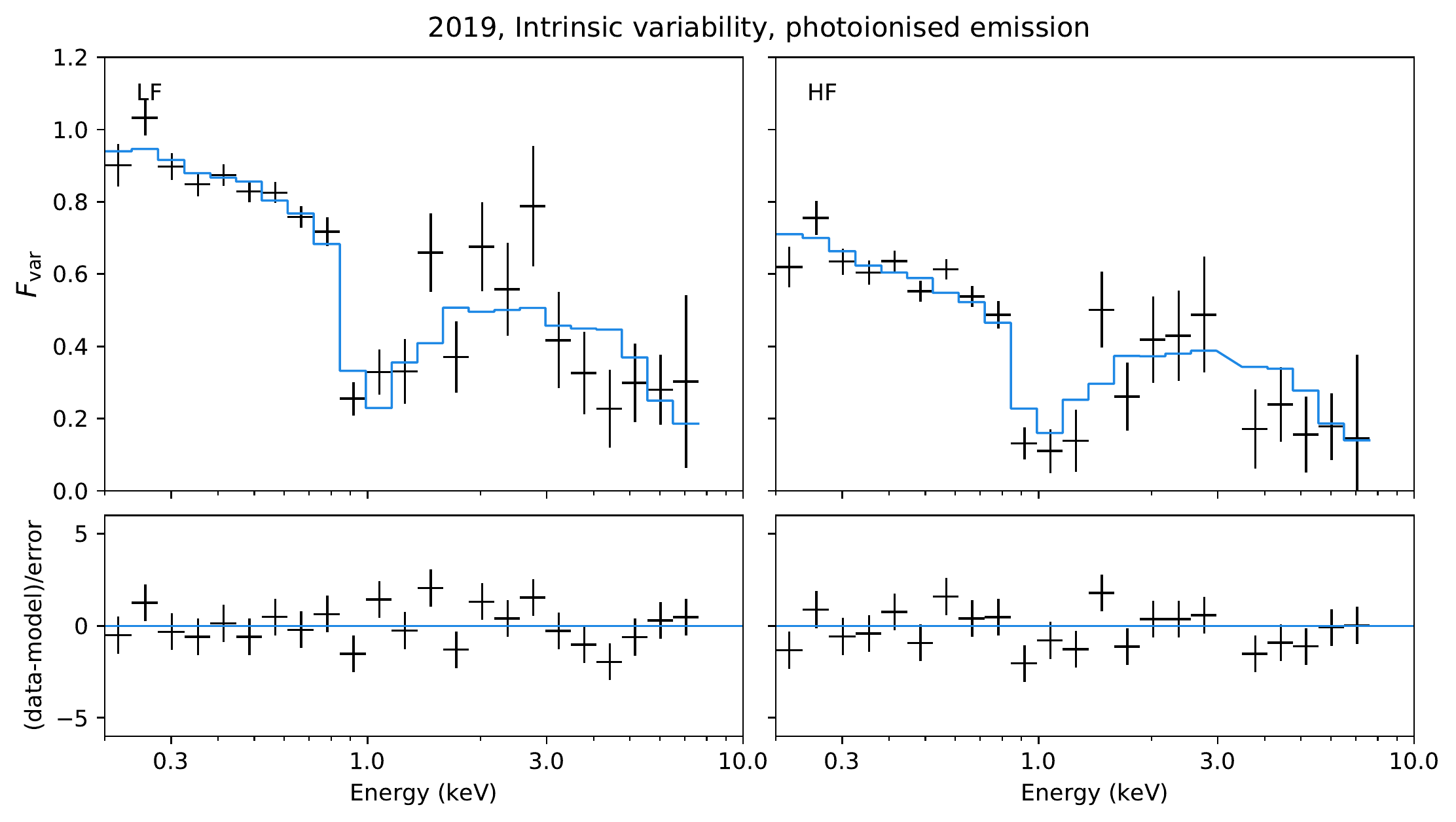}
    \includegraphics[width=0.67\linewidth]{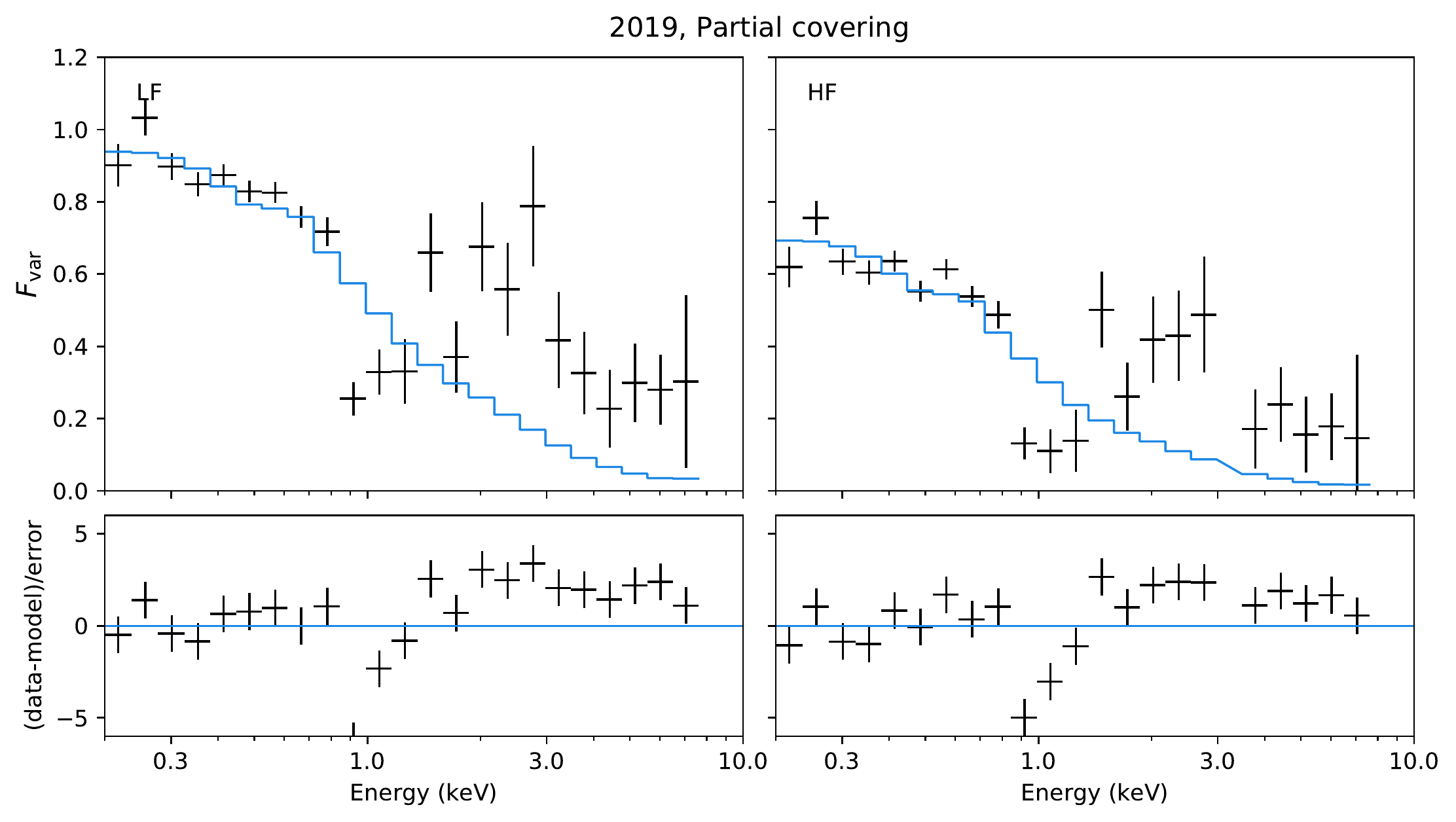}
    \caption{Additional fits to the 2019 \fvar\ spectra. The top panels show the fits with an intrinsic variability model, damped by photoionised emission. The fit is acceptable, but is qualitatively dramatically different from the other spectra. The bottom panels show a fit where the variance is purely due to absorption, with no other processes contributing. The fit captures the variability below 1 keV well, but overpredicts the variance at 1 keV and underpredicts it at high energies.}
    \label{fig:spectra_2019_2}
\end{figure*}


\bsp	
\label{lastpage}
\end{document}